\RequirePackage{lineno}
\documentclass[aps,prd,twocolumn,letterpaper,superscriptaddress, showpacs, amsmath, amssymb]{revtex4}

\usepackage{graphicx}
\usepackage{amsmath,amssymb}
\usepackage{color}

\begin{document}


\title{Constraining high-energy cosmic neutrino sources: Implications and prospects}

\author{Kohta Murase}
\affiliation{Department of Physics; Department of Astronomy \& Astrophysics; Center for Particle and Gravitational Astrophysics, The Pennsylvania State University, University Park, Pennsylvania 16802, USA}
\affiliation{Institute for Advanced Study, Princeton, New Jersey 08540, USA}
\author{Eli Waxman}
\affiliation{Particle Physics \& Astrophysics Dept., Weizmann Institute of Science, Rehovot 76100, Israel}

\date{Submitted 7 July 2015}

\begin{abstract}
We consider limits on the local ($z=0$) density ($n_0$) of extragalactic neutrino sources set by the nondetection of steady high-energy neutrino sources producing $\gtrsim50$~TeV muon multiplets in the present IceCube data, taking into account the redshift evolution, luminosity function and neutrino spectrum of the sources.  We show that the lower limit depends moderately on source spectra and strongly on redshift evolution.  We find $n_0\gtrsim{10}^{-8}-{10}^{-7}~{\rm Mpc}^{-3}$ for standard candle sources evolving rapidly, $n_s\propto{(1+z)}^3$, and $n_0\gtrsim{10}^{-6}-{10}^{-5}~{\rm Mpc}^{-3}$ for nonevolving sources.  The corresponding upper limits on their neutrino luminosity are $L_{{\nu_\mu}}^{\rm eff}\lesssim10^{42}-10^{43}~{\rm erg}~{\rm s}^{-1}$ and $L_{{\nu_\mu}}^{\rm eff}\lesssim10^{41}-10^{42}~{\rm erg}~{\rm s}^{-1}$, respectively.  Applying these results to a wide range of classes of potential sources, we show that powerful ``blazar'' jets associated with active galactic nuclei are unlikely to be the dominant sources.  For almost all other steady candidate source classes (including starbursts, radio galaxies, and galaxy clusters and groups), an order of magnitude increase in the detector sensitivity at $\sim0.1-1$~PeV will enable a detection (as point sources) of the few brightest objects.  Such an increase, which may be provided by next-generation detectors like {\it IceCube-Gen2} and an upgraded KM3NET, can improve the limit on $n_0$ by more than two orders of magnitude.  Future gamma-ray observations (by {\it Fermi}, HAWC and CTA) will play a key role in confirming the association of the neutrinos with their sources.
\end{abstract}

\pacs{95.85.Ry, 98.70.Sa, 98.70.Vc\vspace{-0.3cm}}
\maketitle

\section{Introduction}
The detection of an extraterrestrial high-energy, $\sim30$~TeV to a few PeV, neutrino flux by the IceCube Collaboration~\citep{Aartsen:2013bka,Aartsen:2013jdh,Aartsen:2014gkd,Aartsen:2014muf,Aartsen:2015ita,Aartsen:2015rwa} marks the beginning of high-energy neutrino astrophysics.  The observed signal is consistent with an isotropic arrival distribution of the neutrinos, and with equal contents of $\nu_e$, $\nu_\mu$ and $\nu_\tau$ and their anti-particles.  Above $\sim100$~TeV, the flux and spectrum are consistent with the Waxman-Bahcall (WB) bound~\citep{Waxman:1998yy}, $E_{\nu}^2\Phi_{\nu_i}\simeq{10}^{-8}~{\rm GeV}~{\rm cm}^{-2}~{\rm s}^{-1}~{\rm sr}^{-1}$ per flavor, with a possible spectral break or cutoff at a few PeV.  These properties together hint to a cosmological origin of the observed neutrino flux, most likely related to the accelerators of high-energy cosmic rays (CRs) (see Refs.~\cite{Waxman:2013zda,Halzen:2013dva,Meszaros:2014tta} for reviews).  High-energy neutrinos are expected to be emitted in this case mainly by the decay of mesons and muons produced in interactions of CRs with ambient gas (nucleons) or radiation fields within or surrounding the CR sources, with neutrino to CR energy ratio typically given by $E_\nu/E_{\rm cr}\approx(0.03-0.05)/A$, where $A$ is the CR atomic number~\cite{Waxman:2013zda,Murase:2010gj}.

\begin{figure}[t]
\includegraphics[width=3.00in]{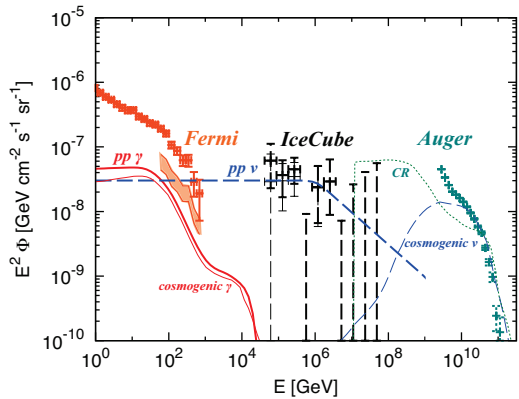}
\caption{
Diffuse CR (thin dotted line), gamma-ray (thick solid line, adapted from Ref.~\cite{Murase:2013rfa}), and all-flavor neutrino (thick dashed line, adapted from Ref.~\cite{Murase:2013rfa}) intensities predicted in our grand-unified cosmic particle model in which the UHECR flux is produced by an extragalactic distribution of proton sources, producing a ``flat" CR proton spectrum, $E_{\rm cr}Q_{E_{\rm cr}}=0.5\times10^{44}~{\rm erg}~{\rm Mpc}^{-3}~{\rm yr}^{-1}$, and residing in environments which are almost ``calorimetric" for $E_{\rm cr}\lesssim50-100$~PeV protons.  The observed UHECR flux and spectrum (Auger data points from Ref.~\cite{ThePierreAuger:2015rha,Aab:2015bza}) and IceCube's neutrino flux and spectrum (IceCube data points from Ref.~\cite{Aartsen:2014gkd}) are both self-consistently explained (see Refs.~\cite{Katz:2013ooa,Waxman:2013zda} for detailed discussion).
The non-blazar contribution of the diffuse gamma-ray background measured by {\it Fermi} (shaded region above 50~GeV), which amounts to $\sim30\%$~\cite{Lisanti:2016jub} (see also Ref.~\cite{TheFermi-LAT:2015ykq}) of the ``total'' extragalactic gamma-ray background, shown as {\it Fermi} data points~\cite{Ackermann:2014usa}, is simultaneously accounted for in this model (see Refs.~\cite{Murase:2013rfa,Murase:2015xka} for details).  The model UHECR flux (thin dotted line) and corresponding cosmogenic neutrino (thin dashed line) and gamma-ray (thin solid line) fluxes are adapted from Ref.~\cite{Decerprit:2011qe}.
\label{Unified}
}
\vspace{-1.\baselineskip}
\end{figure}

IceCube's analysis of lower-energy neutrino events indicates an excess of events at $\sim30$~TeV above an extension to low-energy of the ``flat", $E_{\nu}^2\Phi_{\nu_i}=Const.$, higher-energy spectrum~\cite{Aartsen:2014muf,Aartsen:2015ita}.  Assuming that the astrophysical neutrino spectrum is described by a single power-law, $E_{\nu}^2\Phi_{\nu_i}\propto E_{\nu}^{2-s}$, different analyses of IceCube's data lead to different constraints on the spectral index $s$.
There is some ($\sim2\sigma$) tension between analyses with higher-energy thresholds, yielding values consistent with $s=2$, and those with lower-energy thresholds, yielding $s\sim2.5$ (see Refs.~\cite{Aartsen:2014muf,Aartsen:2015ita,Aartsen:2015zva}).
This may indicate a new component contributing to the flux at $\lesssim100$~TeV energies (see Refs.~\cite{Murase:2015xka,Chen:2014gxa,Palladino:2016zoe,Neronov:2016bnp} for discussion).
The existence of such a component does not affect the analysis presented in this work, which is focused on the higher-energy, $\gtrsim100$~TeV, neutrinos, the flux and spectrum of which are consistent with the WB bound.
It should be noted in this context, that the observed neutrino flux is comparable to the sub-TeV diffuse gamma-ray background flux measured by {\it Fermi}~\cite{Ackermann:2014usa}.  An extension with $s\gtrsim2.1-2.2$ of the $\gtrsim100$~TeV neutrino flux to low energies, $\sim0.1-1$~TeV, implies a diffuse gamma-ray flux that exceeds the {\it Fermi} gamma-ray background, while an extension with a smaller spectral index is consistent with the {\it Fermi} data~\cite{Murase:2013rfa} (see also Fig.~\ref{Unified}).

The coincidence of the IceCube signal with the WB bound implies that the universal average of the energy production rate of ultrahigh-energy (UHE), $>10^{19}$~eV, CRs is similar to the rate of energy production of $\sim0.1-1$~PeV neutrinos.  The observed neutrino signal may thus be explained by a model in which the sources of UHECRs produce protons with a ``flat" spectrum (equal energy per logarithmic particle energy interval), $E_{\rm cr}Q_{E_{\rm cr}}=E_{\rm cr}^2d\dot{n}_{{\rm cr}}/dE_{\rm cr}=Const.$, and reside in ``calorimetric" environments in which protons of energy $\lesssim50-100$~PeV lose all their energy to meson production (i.e. these CRs are confined for a time longer than their $pp$ energy-loss time).
This is the simplest explanation in the sense that the CR sources are known to exist, the required CR spectrum is consistent with that observed at $\gtrsim10^{19}$~eV and with theoretical expectations, the model contains no free parameters (the production rate of CRs is determined by observations and the fraction of their energy converted to mesons is ${\rm min}[f_{pp},1]\simeq1$ below $50-100$~PeV), and there is a known class of objects, which are expected to act as ``calorimeters" for $\lesssim50-100$~PeV protons -- starburst galaxies (SBGs).  In fact, the signal detected by IceCube has been predicted to be produced by sources residing in SBGs~\citep{Loeb:2006tw}.  The only assumption that one needs to make is that CR production is related to star-formation activity (which would be the case for sources like gamma-ray bursts (GRBs), energetic supernovae, or perhaps stellar tidal disruptions by supermassive black holes).  The main uncertainty in this model (see Ref.~\cite{Waxman:2013zda} for a detailed discussion) is related to the fact that galaxies rapidly forming stars are inferred to act as calorimeters for CR protons based on the observations of local ($z=0$) SBGs, while most of the neutrinos are produced by galaxies rapidly forming stars at redshifts $z\sim1-2$.  The properties of these galaxies are less well-constrained, and hence the fraction of them which are ``calorimetric" is uncertain.

While the above unified scenario for the production of UHECRs and of IceCube's neutrinos is simple and natural~\cite{Katz:2013ooa,Waxman:2013zda}, we have no direct confirmation for the emission of neutrinos from SBGs.  A wide range of different models have been proposed for the origin of IceCube's neutrinos.

Models predicting the production of high-energy neutrinos through the decay of mesons and muons produced by high-energy CRs may be divided into two types: ``CR accelerator models", where neutrinos are produced within the CR source, and ``CR reservoir models", where neutrinos are produced while they are confined within the environment surrounding the CR source.
CR accelerator models for the emission of high-energy neutrinos have been proposed, for example, for gamma-ray bursts (GRBs, e.g., Refs.~\cite{Waxman:1997ti,Murase:2006mm,Bustamante:2014oka}) and blazars~\cite{Mannheim:1995mm,Atoyan:2001ey,Atoyan:2002gu,Dermer:2012rg}, while CR reservoir models for the emission of high-energy neutrinos have been proposed for SBGs~\citep{Loeb:2006tw}, galaxy clusters and groups (GCs/GGs)~\cite{Murase:2008yt,Kotera:2009ms}, and active galactic nuclei (AGN)~\cite{Kimura:2014jba,Hooper:2016jls}.
In accelerator models, mesons are typically produced by interactions of CRs with radiation, while in reservoir models they are typically produced by inelastic hadronuclear collisions.  Some AGN core models, where protons are accelerated and undergo hadronuclear collisions in the vicinity of the black hole (e.g., Refs.~\cite{Tjus:2014dna,Kimura:2014jba}), are an exception.

In models where the mesons are produced by photohadronic ($p\gamma$) interactions with radiation, a low-energy cutoff may be expected in the neutrino spectrum.  For a characteristic energy $E_\gamma$ of the ambient photons, the low-energy cutoff is expected at $\sim0.05 E_{\rm min}$, where $E_{\rm min}\sim m_p m_\pi c^4/E_\gamma$ is the minimum CR nucleon energy required to allow pion production (in case the source is moving relativistically with Lorentz factor $\Gamma$, $E_{\rm min}\sim\Gamma^2 m_p m_\pi c^4/E_\gamma$).
In models where mesons are produced by inelastic hadronuclear ($pp$) collisions, we expect the neutrino spectrum to extend down to sub-GeV energies, since pion production is allowed for all relativistic CRs.  In this case, the neutrino spectral index should satisfy $s\lesssim2.1-2.2$, since, as explained above, for steeper spectra the accompanying gamma-ray flux will be inconsistent with the {\it Fermi} gamma-ray background below $\sim0.1-1$~TeV~\cite{Murase:2013rfa}.
If IceCube's neutrinos are produced by $pp$ interactions, the sources significantly contribute to the extragalactic gamma-ray background. This is not necessarily the case for $p\gamma$ scenarios~\cite{Murase:2015xka}, since the radiation field required to produce sub-PeV neutrinos via $p\gamma$ interactions naturally leads to a large two-photon annihilation optical depth for GeV-TeV gamma rays~\cite{Waxman:1998yy,Murase:2015xka}.

\begin{figure}[t]
\includegraphics[width=3.00in]{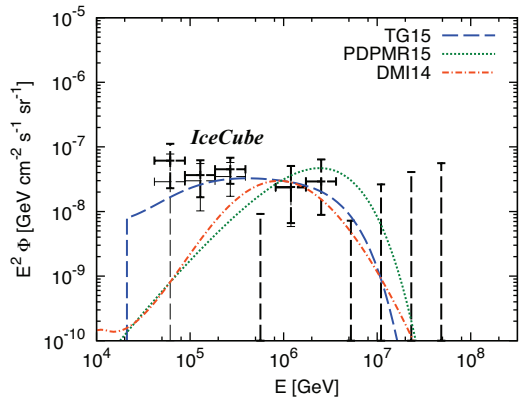}
\caption{All-flavor neutrino fluxes of ``post-IceCube" blazar models, with parameters chosen to explain the IceCube data.  We consider in this paper three spectral templates, taken from Tavecchio \& Ghisellini (TG15)~\cite{Tavecchio:2014eia} and Petropoulou et al. (PDPMR15)~\cite{Petropoulou:2015upa} for BL Lac objects, and from Dermer et al. (DMI14)~\cite{Dermer:2014vaa} for flat spectrum radio sources (FSRQs).
\label{Blazar}
}
\vspace{-1.\baselineskip}
\end{figure}

The neutrino signal detected in IceCube is not consistent with the predictions of most CR accelerator models derived prior to the IceCube detection. Neutrino production within GRB sources is expected to produce a flux which is $\sim10(E_\nu/1~{\rm PeV})$\% of the WB flux at $E_\nu\lesssim1$~PeV~\cite{Waxman:1997ti}.  The neutrino spectra predicted to be produced in AGN jet models (in particular blazar models) are typically inconsistent with (too hard compared to) IceCube's data~\cite{Murase:2014foa,Padovani:2015mba,Murase:2015ndr}.  Nevertheless, the blazar models are not ruled out since their underlying assumptions (e.g., the maximum CR energy) may be modified.  In fact, many of them have been revised after IceCube's discovery, with parameters appropriately chosen to reproduce IceCube's flux above $\sim100$~TeV; in particular see Refs.~\cite{Dermer:2014vaa,Tavecchio:2014xha,Tavecchio:2014eia,Petropoulou:2015upa} and Fig.~\ref{Blazar} for blazar models.
In addition, models of CR accelerators obscured in gamma rays are considered (see Refs.~\cite{Murase:2015xka,Winter:2013cla,Kistler:2013my} and references therein), and
AGN core models~\cite{Stecker:1991vm,Stecker:2013fxa,Kimura:2014jba,Kalashev:2015cma} have been modified such that their flux normalization is adjusted to IceCube's flux. Finally, we note that ``choked jet supernova" models~\cite{Murase:2013ffa,Nakar:2015tma,Senno:2015tsn,Tamborra:2015fzv,Meszaros:2001ms} may also account for the IceCube data.  However, in our current analysis we derive constraints on steady sources and therefore do not address these transient models further.

The main goal of this paper is to demonstrate that the limits that can be set by IceCube's measurements on the source density exclude some widely discussed candidate sources, and to show that an order of magnitude increase in the detector sensitivity at $\sim100$~TeV is likely to enable the detection (as point sources) of the few brightest objects for almost all other candidate source classes.  The limit on the density of ``standard candle" sources is derived in Sec.~\ref{nuA}.  Its implications to various classes of sources are described in Sec.~\ref{nuB}, taking into account the redshift evolution and the luminosity function (LF) of the sources.  The increase in the detector sensitivity required to enable the detection of neutrino point sources (sources producing multiple neutrino events) is discussed in Sec.~\ref{multi1}.

The nondetection of point sources has been used in earlier work \cite{Lipari:2008zf,Silvestri:2009xb,Murase:2012df} to set limits on the density of neutrino sources.  The limits derived here are more stringent thanks to the completion of the full IceCube detector, as recently discussed~\cite{Ahlers:2014ioa,Kowalski:2014zda,Murase:2014JSI,Murase:2015ipa}.  Moreover, our analysis goes beyond those of earlier work in taking into consideration the dependence on the redshift evolution and on the LF of the sources, and also on the neutrino spectrum of the sources.  As explained below, the density limit is sensitive to the redshift evolution, and taking into account the LF of the sources implies that their ``effective" number density $n_0^{\rm eff}$ (the number density of sources dominating the flux), which is constrained by the derived density limit, may be significantly smaller than the total density, $n_0^{\rm tot}$.

The non-blazar component of the sub-TeV extragalactic gamma-ray background flux measured by {\it Fermi}~\cite{TheFermi-LAT:2015ykq,Lisanti:2016jub} can be explained by the sum of hadronic gamma rays produced inside the sources and ``cosmogenic" gamma rays produced in CR interactions with the cosmic microwave background and extragalactic background light (see Fig.~\ref{Unified}).
In particular, SBGs have been predicted to produce a significant contribution to the diffuse gamma-ray background~\cite{Thompson:2006np,Lacki:2010vs,Lacki:2012si}, consistent with the neutrino flux measured by IceCube.
The source density and luminosity reached by gamma-ray observations are discussed in Sec.~\ref{multi2}, where we show that some neutrino source models like CR reservoir models should be testable with future gamma-ray observatories.

Our conclusions are summarized and discussed in Sec.~\ref{sec:discussion}. We use $\Omega_m=0.3$, $\Omega_\Lambda=0.7$ and $H_0=70~{\rm km~s^{-1}~{Mpc}^{-1}}$ throughout.


\section{Source density limits}\label{nuA}
The analysis presented here relies on medium-energy muon-neutrino-induced muon track events, for which the angular resolution ($\sim0.5$~deg) enables one to straightforwardly determine the absence of sources producing multiple events. 
Although statistics are limited, in the high-energy data sets, where the atmospheric backgrounds are much smaller, significant clustering has not been seen in both the latest high-energy starting event (HESE) data (including several tracks)~\cite{Aartsen:2014gkd} and the multiyear upgoing muon neutrino data~\cite{Aartsen:2015rwa,Aartsen:2016xlq} (cf.~\footnote{Two muon events with RA=235~deg and RA=238~deg are relatively close to each other~\cite{Aartsen:2015rwa}.}).  More statistics are available by including lower-energy events, and no source has been detected in the point and extended source analyses~\cite{Aartsen:2014ivk,Aartsen:2014cva,Aartsen:2015zva}. 
As taken into account in the point-source analyses, low-energy doublets may come from the atmospheric neutrino background.  In what follows we consider the implications of a nondetection of any medium- or high-energy multiplets in the multiyear observation by IceCube and the future neutrino detector {\it IceCube-Gen2}.  The background-induced false number of sources producing multiplets is small enough for sufficiently high-energy muon tracks.

We consider in this section the limits set on the number density and luminosity of ``standard candle" sources, all producing the same luminosity. We denote the density and luminosity of the sources by $n_s^{\rm eff}$ and $L_{\nu_\mu}^{\rm eff}$, and explain in Sec.~{\ref{nuB}} how these effective density and luminosity may be defined for nonstandard candle sources in order to enable the application of the results to such source classes. The muon neutrino luminosity is defined as the luminosity per logarithmic neutrino energy bin ($E_\nu L_{E_{\nu_\mu}}\equiv E_\nu dL_{\nu_\mu}/dE_\nu$).

The average (over randomly distributed observers) number of the sources producing more than $k-1$ multiple events is given by $N_{m\geq k}=\int d{\mathcal V}\,n_s^{\rm eff}[z]P_{m\geq k}[z]$, where $P_{m\geq k}[z]$ is the probability that a single source at redshift $z$ will produce more than $k-1$ multiple events, and $n_s^{\rm eff}[z]$ is the comoving source density at $z$ (we assume that, for random observers, the number of sources within small volumes $d{\mathcal V}$ follows a Poisson distribution with average $n_s^{\rm eff}[z]d{\mathcal V}$).  When the number of total signal events is not too small, denoting the average number of events produced by a source at $z$ by $\lambda[z]$, we have $P_{m\geq2}(\lambda)=1-(1+\lambda)\exp(-\lambda)$, where $\lambda$ may be expressed using the luminosity distance $d_{{\mathcal N}=1}$ for which a source produces one event, $\lambda[z]=(d_{{\mathcal N}=1}/d_L[z])^2$ where $d_L[z]$ is the luminosity distance to $z$.

Using the above definitions, the average number of sources producing multiple events may be written as
\begin{eqnarray}
\label{eq:Nm}
N_{m\geq2}&=&n_0^{\rm eff}\Delta\Omega\int dz\,\frac{(c/H_0)d^2_L[z]}{(1+z)^2\sqrt{\Omega_m(1+z)^3+\Omega_\Lambda}}\nonumber\\
&\times&\left(\frac{n_s^{\rm eff}[z]}{n_0^{\rm eff}}\right)P_{m\geq2}(\lambda[z]),
\end{eqnarray}
where $\Delta\Omega$ is the solid angle covered by the detector and $n_0^{\rm eff}=n_s^{\rm eff}[z=0]$ is the local source density.  Note that Eq.~(\ref{eq:Nm}) itself does not assume any connection to the gamma-ray luminosity, and all information on the exposure of neutrino detectors is included through the definition of $d_{{\mathcal N}=1}$ (see Eq.~\ref{eq:dm} and Appendix A).

The above equation is useful when the contribution of the background is negligible. If the distance out to which sources may be identified as producing multiple events is sharply limited by the background as $d_L<d_{\rm lim}$ (i.e. $E_\nu F_{E_{\nu_\mu}}>F_{\rm lim}$), the number of point sources is~\cite{Murase:2012df}
\begin{eqnarray}
\label{eq:Nm2}
N_{\rm lim}&=&n_0^{\rm eff}\Delta\Omega\int^{z_{\rm lim}}dz\,\frac{(c/H_0)d^2_L[z]}{(1+z)^2\sqrt{\Omega_m(1+z)^3+\Omega_\Lambda}}\nonumber\\
&\times&\left(\frac{n_s^{\rm eff}[z]}{n_0^{\rm eff}}\right),
\end{eqnarray}
where $z_{\rm lim}$ is the redshift corresponding to $d_{\rm lim}$.

For $d_{\mathcal N=1}\ll c/H_0$, $N_{m\geq2}$ is given by $N_{m\geq2}\approx\sqrt{\pi}(\Delta\Omega/3)n_0^{\rm eff}d_{\mathcal N=1}^3$.  Similarly, for higher multiplets, we have $N_{m\geq3}\approx\sqrt{\pi/16}(\Delta\Omega/3)n_0^{\rm eff}d_{\mathcal N=1}^3$ (for triplets and higher) and $N_{m\geq4}\approx\sqrt{\pi/64}(\Delta\Omega/3)n_0^{\rm eff}d_{\mathcal N=1}^3$ (for quartets and higher), respectively.  With Eq.~(\ref{eq:Nm2}), we reproduce the well-known result, $N_{\rm lim}\approx(\Delta\Omega/3)n_0^{\rm eff}d_{\rm lim}^3$~\cite{Murase:2012df}.  The calculation does not necessarily rely on the high-energy muon events above $\sim200$~TeV.  One can use the point-source sensitivity that is derived from the track data with more statistics~\cite{Aartsen:2014cva,Aartsen:2015zva} (see also Ref.~\cite{Silvestri:2009xb}).   

For the purpose of placing limits on the source density, we consider $m\geq2$ multiplets.  We therefore write
\begin{equation}\label{eq:Nm_approx}
N_{m\geq2}=\sqrt{\pi}q_{L}\left(\frac{\Delta\Omega}{3}\right)n_0^{\rm eff}d_{\mathcal N=1}^3,
\end{equation}
where the luminosity dependent function $q_L$ depends on redshift evolution models, and approaches unity at sufficiently low luminosities.
For example, for luminosity corresponding to $d_{\mathcal N=1}/(c/H_0)=0.1$ we find $q_L=0.94$ and $q_L=2.0$ for redshift evolution of the form $n_s[z]\propto(1+z)^{m}$ with $m=0$ (no evolution) and $m=3$ (rapid evolution that reasonably mimics the star-formation rate (SFR) or AGN luminosity density), respectively.

For neutrino sources with a ``flat", $E_\nu F_{E_{\nu_\mu}}=Const.$, spectrum in the $0.1-1$~PeV range, the nondetection of point and extended sources in the four-year data of IceCube sets a 90\% CL upper limit of $E_\nu F_{E_{\nu_\mu}}<F_{\rm lim}\approx{10}^{-9}~{\rm GeV}~{\rm cm}^{-2}~{\rm s}^{-1}$ to the {\it muon neutrino} flux produced by a possible point source~\cite{Aartsen:2014cva}.  The two-year sensitivity is worse by a factor of two, while the sensitivity is improved to $F_{\rm lim}\approx(6-7)\times{10}^{-10}~{\rm GeV}~{\rm cm}^{-2}~{\rm s}^{-1}$ with the six-year data~\cite{Aartsen:2015yva}. For the high-energy IceCube data, that are essentially background free, a 90\% CL upper limit corresponds to an upper limit of $\lambda<2.44$ on the number of events produced on average by a source~\cite{Feldman:1997qc}. Denoting the source differential ``muon neutrino" luminosity by $L_{E_{\nu_\mu}}^{\rm eff}=d L^{\rm eff}_{\nu_\mu}/dE_\nu $, we have
\begin{eqnarray}\label{eq:dm}
d_{\mathcal N=1}&\approx&\left(\frac{E_\nu L_{E_{\nu_\mu}}^{\rm eff}}{4\pi F_{\rm lim}/2.4}\right)^{1/2}\nonumber\\
&\simeq&110~{\rm Mpc}~\left(\frac{E_\nu L_{E_{\nu_\mu}}^{\rm eff}}{{10}^{42}{\rm erg\,s^{-1}}}\right)^{1/2}F_{\rm lim, -9}^{-1/2},
\end{eqnarray}
where $F_{\rm lim}={10}^{-9}F_{\rm lim, -9}~{\rm GeV}~{\rm cm}^{-2}~{\rm s}^{-1}$. 

Interpreting the absence of multiple event sources as a limit on $N_{m\geq k}$, we may impose $N_{m\geq k}<1$ (or $N_{m\geq k}/N_b<1$ in the presence of significant backgrounds, where $N_b$ is the number of false multiplet sources).  In more general, one may write the condition as
\begin{equation}\label{eq:Nm_general}
\hat{N}_s=b_{m,L}\left(\frac{\Delta\Omega}{3}\right)n_0^{\rm eff}d_{\rm lim}^3<1\nonumber,
\end{equation}
where $b_{m,L}$ is an order-of-unity factor that depends on details of analyses. For example, if we consider $m\geq2$ multiplets and $N_b\lesssim1$ (that is satisfied for the assumed threshold and exposure), we obtain $b_{m,L}\simeq6.6q_L$.  Note that Eq.~(\ref{eq:Nm}) gives a stronger limit that from Eq.~(\ref{eq:Nm2}), as seen from $b_{m}>1$.  This is because there is a nonnegligible contribution of distant neutrino sources (from $z>z_{\rm lim}$) to doublet sources, due to $P_{m\geq2}(\lambda)$. On the other hand, as naturally expected, higher-multiplet sources are more largely contributed by nearby neutrino sources. Indeed, for triplets or higher multiplets, we obtain $b_{m}\simeq1.6$.  
 
Using Eqs.~(\ref{eq:Nm_approx}) and (\ref{eq:dm}), the condition $N_{m\geq2}<1$ gives
\begin{eqnarray}\label{eq:n0a}
n_0^{\rm eff}\left(\frac{E_\nu L_{E_{\nu_\mu}}^{\rm eff}}{{10}^{42}{\rm erg\,s^{-1}}}\right)^{3/2}F_{\rm lim, -9}^{-3/2}&&\nonumber\\
\lesssim1.9\times10^{-7}&{\rm Mpc^{-3}}&q_L^{-1}\left(\frac{2\pi}{\Delta\Omega}\right).
\end{eqnarray}
Note that this gives an {\it upper limit on $n_0^{\rm eff}$}, which depends on the luminosity (consistent with the results of Refs.~\cite{Silvestri:2009xb,Murase:2012df,Waxman:2008bj}, in contrast with the result of Ref.~\cite{Kowalski:2014zda}). The upper limit is insensitive to the redshift evolution at sufficiently low luminosities, and is valid regardless of whether or not the sources dominate IceCube's neutrino flux.

The diffuse neutrino intensity observed by IceCube determines the neutrino luminosity density of the Universe, $n_0^{\rm eff}(E_\nu L_{E_{\nu_\mu}}^{\rm eff})$. The coincidence of the observed intensity with the WB flux enables one to determine the neutrino luminosity density by using Eqs.~(1), (2) and (5) of Ref.~\cite{Waxman:1998yy}, from which we find
\begin{eqnarray}\label{eq:n0L}
n_0^{\rm eff}\left(\frac{E_\nu L_{E_{\nu_\mu}}^{\rm eff}}{10^{42}{\rm erg\,s^{-1}}}\right)&\simeq&1.6\times{10}^{-7}~{\rm Mpc^{-3}}~(3/\xi_z)\nonumber\\
&\times&\left(\frac{E_\nu^2 \Phi_{\nu_\mu}}{10^{-8}~{\rm GeV}~{\rm cm}^{-2}~{\rm s}^{-1}~{\rm sr}^{-1}}\right),\,\,\,\,\,\,\,\,\,\,
\end{eqnarray}
where $\xi_z$ is a dimensionless parameter that depends on the redshift evolution of the sources: $\xi_z\approx3$ for $m=3$ and $\xi_z\approx0.6$ for $m=0$~\cite{Waxman:1998yy} ($\xi_z\approx2.8$ for SFR evolution~\cite{Hopkins:2006bw}, $\xi_z\approx8.4$ for FSRQ evolution, and $\xi_z\approx0.68$ for BL Lac evolution~\cite{Ajello:2013lka}). Combining Eqs.~(\ref{eq:n0a}) and (\ref{eq:n0L}), we find
\begin{equation}\label{eq:Lnu}
\left(\frac{E_\nu L_{E_{\nu_\mu}}^{\rm eff}}{10^{42}{\rm erg\,s^{-1}}}\right)\lesssim1.4~q_L^{-2}\left(\frac{\xi_z}{3}\right)^{2}F^{3}_{\rm lim, -9}\left(\frac{\Delta \Omega}{2\pi}\right)^{-2},
\end{equation}
and
\begin{equation}\label{eq:n0}
n_0^{\rm eff}\gtrsim1.1\times10^{-7}\,{\rm Mpc^{-3}}\,q_L^2\left(\frac{\xi_z}{3}\right)^{-3}F^{-3}_{\rm lim, -9}\left(\frac{\Delta \Omega}{2\pi}\right)^2.
\end{equation}
Note that Eq.~(\ref{eq:n0}) gives a {\it lower limit}, which can be placed because we {\it require} that the considered standard candle sources produce the neutrino flux detected by IceCube.
Remarkably, the constraints are quite sensitive to the redshift evolution, and are more stringent for weaker evolution.  This is simply because $\xi_z$ in Eq.~(\ref{eq:n0L}) comes via the cubic term in Eq.~(\ref{eq:Nm_approx}).  The background becomes more important at lower energies or longer exposure time or poorer angular resolution. If the false number of multiplet sources is $N_b\sim2-3$, the lower limit is relaxed by a factor of $4-9$.  Instead, if Eq.~(\ref{eq:Nm2}) is used or $m\geq3$ multiplets are considered more conservatively, the lower limit changes by a factor of $\sim10$.  Also, its precise value might be affected by details of the muon neutrino data because of its dependence on $F_{\rm lim}$ (that slightly varies with zenith angle).  However, in either case, our discussion on implications and prospects is unaltered.

\begin{figure}[t]
\includegraphics[width=3.00in]{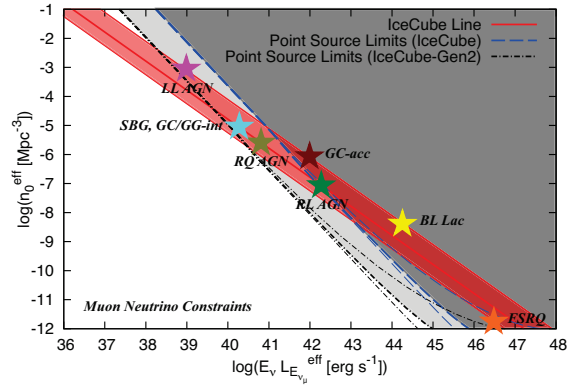}
\caption{IceCube's detection determines the local ($z=0$) neutrino emissivity of the Universe, $n_0^{\rm eff}E_\nu L_{E_{\nu_\mu}}^{\rm eff}$, up to uncertainty related to the unknown redshift evolution of the sources (see Eq.~\ref{eq:n0L}).  The solid ``IceCube lines" show the value of $n_0^{\rm eff}E_\nu L_{E_{\nu_\mu}}^{\rm eff}$ implied by observations for no evolution ($n_s\propto(1+z)^0$, top thin), SFR evolution~\cite{Hopkins:2006bw} (similar to $n_s\propto(1+z)^3$ and AGN evolution~\cite{Ueda:2014tma}, middle thick), and rapid FSRQ evolution (bottom thin).  Nondetection of point sources excludes the shaded regions lying to the right of the dashed and dash-dotted lines (see Eq.~\ref{eq:n0a}), corresponding to the sensitivity obtained for a six-year observation period with IceCube (dashed lines) and a ten-year observation period with {\it IceCube-Gen2} (dot-dashed lines).  
Thick dashed and dash-dotted lines are for SFR evolution, whereas thin dashed and dash-dotted lines are for no evolution (upper curves) and FSRQ evolution (lower curves).  
The flat spectrum template shown in Fig.~\ref{Unified} is used.
Colored stars represent the density and luminosity of various classes of candidate sources.
\label{IClim1}
}
\vspace{-1.\baselineskip}
\end{figure}
\begin{figure}[t]
\includegraphics[width=3.00in]{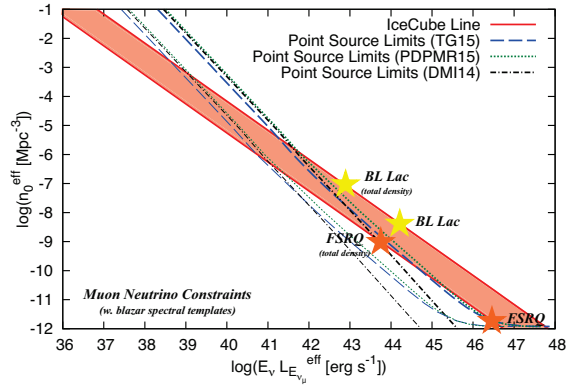}
\caption{Similar to Fig.~\ref{IClim1}, for the blazar spectral templates shown in Fig.~\ref{Blazar}, with BL Lac evolution for the TG15~\cite{Tavecchio:2014eia} and PDPMR15~\cite{Petropoulou:2015upa} templates and FSRQ evolution (with mean redshift $\bar{z}=2$~\cite{Dermer:2014vaa}) for the DMI14 template. The solid red ``IceCube lines" show the value of $n_0^{\rm eff}E_\nu L_{E_{\nu_\mu}}^{\rm eff}$ implied by observations for BL Lac evolution (upper curve) and FSRQ evolution (lower curve). Point source limits obtained for a six-year observation period with IceCube and a ten-year observation period with {\it IceCube-Gen2} are shown with thick and thin lines respectively.
Nondetection of points sources excludes the regions lying to the right and above the dashed, dotted and dash-dotted lines.
\label{IClim2}
}
\vspace{-1.\baselineskip}
\end{figure}

In Fig.~\ref{IClim1} we show the limits obtained using numerical calculations.  In order to estimate the sensitivity, we evaluate the number of through-going muons for both the signal and the background, taking into account the zenith and energy dependence of the effective area of IceCube and the absorption of neutrinos within the Earth (see Appendix A for details). Then, we calculate the probability to find at least one medium- or high-energy multiplet, and place upper limits on $n_0^{\rm eff}$ for different redshift evolution models. The limits obtained numerically are consistent with those obtained analytically above.  For SFR evolution, we find $n_0^{\rm eff}\gtrsim{10}^{-7}~{\rm Mpc}^{-3}$ and $E_\nu L_{E_{\nu_\mu}}^{\rm eff}\lesssim{10}^{42}~{\rm erg}~{\rm s}^{-1}$, consistent with the analytical estimates given by Eqs.~(\ref{eq:n0}) and (\ref{eq:Lnu}).
For $E_\nu L_{E_{\nu_\mu}}^{\rm eff}\sim{10}^{44}~{\rm erg}~{\rm s}^{-1}$, that corresponds to $d_{\mathcal N=1}/(c/H_0)=0.1$, we find $q_L\approx0.9$ for no evolution and $q_L\approx2$ for SFR evolution, consistent with the analytic results.

As seen from Fig.~\ref{IClim1} and Eq.~(\ref{eq:n0}), the lower limit on $n_0^{\rm eff}$ is sensitive to redshift evolution models.  As a result, for nonevolving sources, $m=0$ and $\xi_z\approx0.6$, the limits we can achieve are $n_0^{\rm eff}\gtrsim0.9\times10^{-5}~{\rm Mpc^{-3}}$ and $E_\nu L_{E_{\nu_\mu}}^{\rm eff}\lesssim9\times10^{40}~{\rm erg}~{\rm s}^{-1}$, respectively.  The former (latter) is two orders (one order) of magnitude stronger than the SFR case.
Note that the absence of multiplets in the two-year muon neutrino data (including the public high-energy data set~\cite{Aartsen:2015rwa}) leads to the lower limit of $n_0^{\rm eff}\gtrsim10^{-8}-10^{-7}~{\rm Mpc^{-3}}$ (as in Ref.~\cite{Kowalski:2014zda}), giving an interesting constraint on BL Lac objects (see Section~\ref{nuB}).

The effective area of {\it IceCube-Gen2} is expected to be $\sim5-7$ times larger than IceCube-86~\cite{Aartsen:2014njl}, yielding $F_{\rm lim}\sim{10}^{-10}~{\rm GeV}~{\rm cm}^{-2}~{\rm s}^{-1}$ after $\sim10$~year observations at sufficiently high energies and improving the source density lower limit to $n_0^{\rm eff}\gtrsim0.4\times10^{-4}~{\rm Mpc^{-3}}$ for the $m=3$ or SFR case (see Eq.~\ref{eq:n0}).

The muon neutrino constraints depend not only on redshift evolution models but also on the assumed neutrino spectra, since the limits depend on $F_{\rm lim}$, which in turn is affected by the assumed source spectra.  Although a flat spectrum is a reasonable assumption for CR reservoirs, the neutrino spectrum may be more complicated, as often predicted for blazar models (see Fig.~\ref{Blazar}).  We have expanded our numerical analysis to sources with harder spectra ($s<2$) using the three blazar spectral templates shown in Fig.~\ref{Blazar}, and tested the applicability of Eqs.~(\ref{eq:Lnu}) and (\ref{eq:n0}) for these spectra.  
Our numerical results are shown in Fig.~\ref{IClim2}.  As expected, the limits are somewhat weaker for harder neutrino spectra.

\section{Implications to candidate sources}\label{nuB}
In order to determine the implications of the constraints given by Eqs.~(\ref{eq:Lnu}) and (\ref{eq:n0}) to different classes of candidate neutrino sources, one must take into account the source luminosity distribution (i.e. deviations from ``standard candle" sources).  While the distribution of electromagnetic luminosities, i.e. the photon LFs, of different classes of objects are known, the neutrino LFs of most source classes are not known and are model dependent.  We therefore do not attempt here a comprehensive analysis under different model assumptions regarding the LFs of various classes of objects.  Rather, for each class of objects we define an effective neutrino luminosity, $L^{\rm eff}_{\nu_\mu}$, as the luminosity that maximizes $L_{\nu_\mu}(dn_s/d\ln L_{\rm ph})$ under commonly used model assumptions determining the dependence of $L_{\nu_\mu}$ on the photon luminosity $L_{\rm ph}$, and an effective source number density,
\begin{equation}\label{eq:neff}
n_s^{\rm eff}\equiv\frac{1}{L_{\nu_\mu}^{\rm eff}}\int d (\ln L_{\rm ph})\,L_{\nu_\mu} \frac{dn_s}{d\ln L_{\rm ph}},
\end{equation}
which characterizes the density of sources that dominate the neutrino production.  As we show below, the classes of sources that are ruled out by the constraint of Eq.~(\ref{eq:n0}) are characterized by $n_0^{\rm eff}$ values which are orders of magnitude smaller than the limit of Eq.~(\ref{eq:n0}).

The functional dependence of $L_{\nu_\mu}$ on $L_{\rm ph}$, which is typically of the form $L_{\nu_\mu}\propto L^\alpha_{\rm ph}$, determines $n_0^{\rm eff}$ and $L^{\rm eff}_{\rm ph}$. The absolute value of $L_{\nu_\mu}$, and hence $L_{\nu_\mu}^{\rm eff}$, is typically uncertain and may be considered as a free parameter of the models. It is determined by the requirement that the sources would produce the observed neutrino flux, i.e. by Eq.~(\ref{eq:n0L}). The SBG ``calorimetric" model is an exception--in this model the neutrino luminosity is directly related to the gamma-ray luminosity, and the cumulative flux predicted by the model is consistent with the flux measured by IceCube.

\begin{table}[t]
\begin{center}
\caption{Densities of various classes of steady sources suggested to produce the flux of high-energy neutrinos observed in IceCube.
}
\scalebox{0.82}{
\begin{tabular}{|c|c|c|c|c|}
\hline Source class & $E_\nu L_{E_{\nu_\mu}}^{\rm eff}$ [erg s$^{-1}$] & $L_{\rm ph}^{\rm eff}$ [erg s$^{-1}$] & $n_0^{\rm eff}$ [Mpc$^{-3}$] & $n_0^{\rm tot}$ [Mpc$^{-3}$]\\
\hline ${\rm FSRQ\footnote{Based on the FSRQ LF and redshift evolution of {\it Fermi}~\cite{Ajello:2013lka}; $L_\gamma$ is defined in the [0.1~GeV, 100~GeV] photon energy band.}}$ & $\sim3\times{10}^{46}$ & $L_{\gamma}\sim5\times{10}^{47}$ & $\sim2\times{10}^{-12}$ & $\sim{10}^{-9}$\\
\hline ${\rm BL~Lac}\footnote{Based on the BL Lac LF and redshift evolution of {\it Fermi}~\cite{Ajello:2013lka}; $L_\gamma$ is defined in the [0.1~GeV, 100~GeV] photon energy band.}$ & $\sim2\times{10}^{44}$ & $L_{\gamma}\sim5\times{10}^{45}$ & $\sim5\times{10}^{-9}$ & $\sim{10}^{-7}$\\
\hline ${\rm SBG}$\footnote{Using $L_\gamma\propto L_{\rm IR}^{1.17}$~\cite{Ackermann:2012vca}, where $L_{\rm IR}$ is the infrared luminosity, and the infrared LF of Ref.~\cite{Gruppioni:2013jna}, assuming the SFR redshift evolution (that is similar to the $m=3$ redshift evolution); $L_\gamma$ is defined in the [0.1~GeV, 100~GeV] photon energy band and $L_{\rm IR}$ is defined in the [8~$\mu$m, 1000~$\mu$m] photon energy band.} & $\sim2\times{10}^{40}$ & $L_{\gamma}\sim{10}^{41}$ & $\sim{10}^{-5}$ & $\sim3\times{10}^{-5}$\\
\hline GC-acc\footnote{Based on the halo mass function~\cite{Warren:2005ey}, assuming no redshift evolution; $L_{X}$ is defined in the [0.01~keV, 40~keV] photon energy band.} &$\sim1\times{10}^{42}$ &$L_{X}\sim8\times{10}^{44}$& $\sim{10}^{-6}$ & $\sim2\times{10}^{-6}$\\
GC/GG-int\footnote{Based on the halo mass function~\cite{Warren:2005ey}, assuming the $m=3$ redshift evolution.} & $\sim2\times{10}^{40}$ &$L_{X}\sim6\times{10}^{43}$& $\sim{10}^{-5}$ & $\sim5\times{10}^{-5}$\\
\hline ${\rm RL~AGN}$\footnote{Using $L_\gamma\propto L_{\rm radio}^{1.16}$~\cite{Inoue:2011bm}, where $L_{\rm radio}$ is the radio luminosity, and the radio LF of Ref.~\cite{Willott:2000dh}, assuming  the $m=3$ redshift evolution (that roughly mimics the RL AGN redshift evolution); $L_\gamma$ is defined in the [0.1~GeV, 10~GeV] photon energy band and $L_{\rm radio}$ is defined in the 5~GHz photon energy band.} & $\sim2\times{10}^{42}$ &$L_{\gamma}\sim{10}^{43}$& $\sim{10}^{-7}$ & $\sim{10}^{-4}$\\
\hline ${\rm RQ~AGN}$\footnote{Based on the AGN x-ray LF and redshift evolution of Ref.~\cite{Ueda:2014tma}; $L_{X}$ is defined in the [0.2~keV, 10~keV] photon energy band.} & $\sim7\times{10}^{40}$ &$L_{X}\sim{10}^{44}$& $\sim3\times{10}^{-6}$ & $\sim{10}^{-4}$\\
\hline ${\rm LL~AGN}$\footnote{Based on the H$\alpha$ LF~\cite{Ho:2008rf}, assuming no redshift evolution.} & $\sim1\times{10}^{39}$ &$L_{{\rm H}\alpha}\sim{10}^{40}$& $\sim{10}^{-3}$ & $\gtrsim{10}^{-2}$\\
\hline
\end{tabular}
}
\end{center}
\end{table}

Table~1 presents the values of $n_0^{\rm eff}$ and the corresponding values of $L^{\rm eff}_{\rm ph}\equiv L_{\rm ph}(L^{\rm eff}_{\nu_\mu})$ and $L_{\nu_\mu}^{\rm eff}$, for commonly discussed source classes. The total number density of the sources, $n_0^{\rm tot}$, which is approximately the density of the lowest-power sources, is also indicated.
As explained in some detail below, comparing the numbers given in Table~1 with the constraints on source density, which were derived in the preceding section, implies that rare sources, such as powerful blazar jets (BL Lac objects and FSRQs), are unlikely to be the dominant sources of IceCube's neutrinos.

\noindent (i) FSRQs: The neutrino emission from FSRQs is expected to be dominated by the decay of pions produced via interactions of high-energy protons with external target photons provided by the accretion disk, broad-line region, and dust torus~\cite{Atoyan:2001ey,Atoyan:2002gu}.  The broad-line emission and/or the infrared emission from the dust torus are typically dominant in luminous quasars, and the optical and infrared data imply that the photomeson production efficiency $f_{p\gamma}(\lesssim1)$ is proportional to $L_{\rm AD}^{1/2}$~\cite{Murase:2014foa}.  Here $L_{\rm AD}$ is the accretion disk luminosity and we assume that the CR luminosity is proportional to $L_{\rm AD}$.  Using this simple scaling ($L_{\nu_\mu}\propto L_\gamma^{3/2}$) and the FSRQ LF observed by {\it Fermi}~\cite{Ajello:2013lka} one finds $L_{\gamma}^{\rm eff}\sim5\times{10}^{47}~{\rm erg}~{\rm s}^{-1}$ and $n_0^{\rm eff}\sim2\times{10}^{-12}~{\rm Mpc}^{-3}$, well below the IceCube lower limit on the density of sources given by Eq.~(\ref{eq:n0}) and by Fig.~\ref{IClim2}, $n_0^{\rm eff}\gtrsim{10}^{-9}~{\rm Mpc}^{-3}$.  We note that the total number density of FSRQs, $n_0^{\rm tot}\sim{10}^{-9}~{\rm Mpc}^{-3}$, is comparable to the limit on source density, implying that a model in which the neutrino emission is dominated by the lowest-power FSRQs would be consistent with the source density limit. Such a model is, however, theoretically unlikely.

\noindent (ii) BL Lac objects: In BL Lac objects, internal synchrotron photons in the AGN jet are the most important target photons. One has to take into account that the photomeson production efficiency depends on the spectral energy distributions~\cite{Murase:2014foa,Tavecchio:2014eia,Petropoulou:2015upa}, which may vary with the blazar luminosity (the so-called ``blazar sequence'').  As an example, in Table~1, we consider the TG15 model, that predicts approximately $f_{p\gamma}\propto L_{\gamma}$ at the neutrino energies of interest.  Based on the LF of BL Lac objects~\cite{Ajello:2013lka}, one finds $L_{\gamma}^{\rm eff}\sim5\times{10}^{45}~{\rm erg}~{\rm s}^{-1}$ and $n_0^{\rm eff}\sim5\times{10}^{-9}~{\rm Mpc}^{-3}$, well below the lower limit on the density of sources given by Eq.~(\ref{eq:n0}) and by Fig.~\ref{IClim2} for sources with weak redshift evolution characterizing BL Lac objects~\cite{Ajello:2013lka}, $n_0^{\rm eff}\gtrsim(1-4)\times{10}^{-6}~{\rm Mpc}^{-3}$.  Note that these limits are also larger than $n_0^{\rm tot}\sim{10}^{-7}~{\rm Mpc}^{-3}$.

\noindent (iii) SBGs: In the SBG model, we expect the gamma-ray and neutrino luminosities to be linearly correlated~\cite{Loeb:2006tw}.  In Table~1 we use, following Ref.~\cite{Ackermann:2012vca}, $L_{\nu_\mu}\propto L_\gamma\propto L_{\rm IR}^{1.17}$ (that is roughly consistent with the ``calorimetric" picture), where $L_{\rm IR}$ is the infrared luminosity.  Based on the LF and redshift evolution inferred by the {\it Herschel} surveys at the far-infrared band, we find $L_{\rm IR}^{\rm eff}\sim3\times{10}^{11}~L_{\odot}$, $n_0^{\rm eff}\sim{10}^{-5}~{\rm Mpc}^{-3}$ and $E_\nu L_{E_{\nu_\mu}}^{\rm eff}\sim{10}^{41}~{\rm erg}~{\rm s}^{-1}$.  This source density is well above the lower limit placed by IceCube, $n_0^{\rm eff}\gtrsim{10}^{-7}~{\rm Mpc}^{-3}$, but accessible to next-generation neutrino detectors such as {\it IceCube-Gen2} (see Eq.~\ref{eq:n0} and Fig.~\ref{IClim1}).  
As pointed out by Refs.~\cite{Murase:2014foa,Tamborra:2014xia,Wang:2016vbf}, a significant fraction of SBGs may coexist with AGN (that are mostly radio quiet), and CRs may be accelerated by jets embedded in the galaxy or disk-driven outflows. Such an SBG-AGN model may have similar predictions if they have the typical AGN evolution but their redshift evolution could be as fast as the FSRQ one.
As discussed in Sec.~\ref{multi1} and Sec.~\ref{multi2}, the tight relationship between the neutrino and gamma-ray luminosities predicted in the SBG model implies that the model is testable by future neutrino and gamma-ray detectors.

\noindent (iv) GCs/GGs: Two types of models should be considered here.  In the GC-acc model CRs are produced by the accretion shocks in massive clusters and/or by GC merger shocks~\cite{Keshet:2002sw,Kushnir:2009vm}, and the CR production rate is expected to be proportional to $M^{5/3}$, where $M$ is the cluster halo mass.  Ignoring details such as the halo mass dependence of the gas fraction~\cite{Colafrancesco:1998us}, the $pp$ production efficiency $f_{pp}$ is expected to be proportional to $M^0$ (in the confinement limit) or $M^{2/3}$ (in the diffusion limit).  Assuming $f_{pp}\propto M^{1/3}$ on average, $L_{\nu_\mu}\propto M^{2}$ leading to $L_{X}^{\rm eff}\sim{10}^{45}~{\rm erg}~{\rm s}^{-1}$ (using the $L_X-M$ relation of Ref.~\cite{Reiprich:2001zv}) and $n_0^{\rm eff}\sim{10}^{-6}~{\rm Mpc}^{-3}$.  This density is well below the source density limit for the relevant redshift evolution, $n_s\propto(1+z)^m$ with $m=0$ (no evolution) or even $m<0$.  For nonevolving sources $\xi_z\approx0.6$ the IceCube lower limit is $n_0^{\rm eff}\gtrsim{10}^{-5}{\rm Mpc}^{-3}$ (see Eq.~\ref{eq:n0} and Fig.~\ref{IClim1}), implying that if CRs in GCs are produced by accretion and merger shocks, these objects cannot contribute much to the flux of neutrinos detected by IceCube, which is consistent with the previous calculations~\cite{Berezinsky:1996wx,Colafrancesco:1998us,Murase:2008yt,Zandanel:2014pva,Fang:2016amf}.

In the GC/GG-int model CRs are mainly supplied by sources residing with the GC or GG~\cite{Berezinsky:1996wx,Murase:2008yt,Kotera:2009ms}, like AGN, galaxies, and galaxy mergers.
In this case we expect the CR production rate to be proportional to $M$ and more CR accelerators may be active in the past, and the redshift evolution can be positive~\cite{Murase:2013rfa,Zandanel:2014pva}.  Assuming $n_s\propto(1+z)^3$ leads to $L_{X}^{\rm eff}\sim{10}^{44}~{\rm erg}~{\rm s}^{-1}$ and $n_0^{\rm eff}\sim{10}^{-5}~{\rm Mpc}^{-3}$ (for $L_{\nu_\mu}\propto M^{1.5}$).  This source density is well above the lower limit placed by IceCube, $n_0^{\rm eff}\gtrsim{10}^{-7}~{\rm Mpc}^{-3}$, but accessible to next-generation neutrino detectors such as {\it IceCube-Gen2} (see Eq.~\ref{eq:n0} and Fig.~\ref{IClim1}).

\noindent (v) Misaligned radio-loud (RL) AGN: A nearly linear relation between gamma-ray and radio luminosity, $L_\gamma\propto L_{\rm radio}^{1.16}$, has been inferred from a small sample of gamma-ray detected radio galaxies~\cite{Inoue:2011bm}.  Although the process responsible for gamma-ray emission of many radio galaxies is most likely inverse-Compton scattering (as suggested by its variability~\cite{Acero:2015hja}), various models have been suggested where the gamma-ray emission is produced by the decay of mesons and muons (produced in interactions of high-energy protons with surrounding plasma~\cite{Kotera:2009ms,Murase:2013rfa,Pfrommer:2013eoa,Kimura:2014jba,Fujita:2015xva,Hooper:2016jls}).
In such models one would expect $L_{\nu_\mu}\propto L_\gamma\propto L_{\rm radio}^{1.16}$, implying, based on the radio LF~\cite{Willott:2000dh}, $L_{\gamma}^{\rm eff}\sim{10}^{43}~{\rm erg}~{\rm s}^{-1}$ and $n_0^{\rm eff}\sim{10}^{-7}~{\rm Mpc}^{-3}$.
The inferred number density is close to the lower limit set by IceCube for sources following SFR/AGN evolution (see Eq.~\ref{eq:n0} and Fig.~\ref{IClim1}), implying that RL AGN could contribute significantly to the flux detected by IceCube, and that some RL AGN may be soon detected as neutrino point sources, if the flux is indeed dominated by this class of objects.

\noindent (vi) Radio-quiet (RQ) AGN: Most AGN do not have powerful jets, have been suggested as efficient neutrino sources~\cite{Stecker:1991vm,AlvarezMuniz:2004uz}, in which CRs are accelerated to high energies and lose most of their energy to pion production, ${\rm min}[1,f_{p\gamma}]\sim1$, by interactions with radiation in the vicinity of the supermassive black hole.  Although the original predictions of this model are inconsistent with IceCube's flux, the model may be adjusted to explain the IceCube data~\cite{Stecker:2013fxa,Kalashev:2015cma}.  The simple scaling of this model, $L_{\nu_\mu}\propto L_X$, implies $L_{X}^{\rm eff}\sim{10}^{44}~{\rm erg}~{\rm s}^{-1}$ and $n_0^{\rm eff}\sim3\times{10}^{-6}~{\rm Mpc}^{-3}$ (based on the x-ray LF and redshift evolution of Ref.~\cite{Ueda:2014tma}).  This AGN core model is unconstrained by the present IceCube data, but can be tested with {\it IceCube-Gen2}.

\noindent (vii) Low-luminosity (LL) AGN: It has been suggested that CR acceleration followed by $pp$ and $p\gamma$ interactions in radiatively inefficient accretion flows in the vicinity of the black hole, may account for IceCube's neutrino flux~\cite{Kimura:2014jba}. In this model, the diffuse neutrino flux is likely to be dominated by objects with $L_{{\rm H}\alpha}\sim{10}^{40}~{\rm erg}~{\rm s}^{-1}$~\cite{Kimura:2014jba}, implying $n_0^{\rm eff}\sim10^{-3}~{\rm Mpc}^{-3}$.  Regardless of the redshift evolution of LL AGN, which is currently uncertain~\cite{Ho:2008rf}, this model is not constrained by current IceCube data.  However, if their evolution is weak as that of low-luminosity BL Lac objects and Fanaroff-Riley I radio galaxies, Fig.~\ref{IClim1} suggests that this model could be tested by {\it IceCube-Gen2}.

\section{Searching for the brightest neutrino sources: muon neutrinos}\label{multi1}
Several attempts have been made to identify the neutrino sources by searching for a cross correlation between the neutrino arrival directions and the angular locations on the sky of various types of astrophysical objects (see, e.g., Refs.~\cite{Anchordoqui:2014yva,Padovani:2014bha,Sahu:2014fua,Moharana:2015nxa,Emig:2015dma,Moharana:2016mkl}).
Most of them are based on IceCube's HESE searches, where the fully contained events are selected and many of the neutrino events consist of shower events with $\sim(10-20)$~deg angular resolutions.  Although possible associations have been claimed, none of them are significant.

Stacking analyses or cross-correlation studies can be more powerful than searches for event clustering~\cite{Ahlers:2014ioa}.  However, they intrinsically require multimessenger observations, and it is not straightforward to obtain implications to the candidate sources.  This is because one needs to determine the multimessenger relationship between $L_{\nu_\mu}$ and $L_{\rm ph}$.  Despite the uncertainties, Table~1 suggests that all canonical models considered here should be testable by IceCube and {\it IceCube-Gen2}.

First, we discuss blazars including (i) FSRQs and (ii) BL Lac objects, which are disfavored by the muon neutrino constraints described in the previous section.
For FSRQs, the few brightest objects include 3C 273 and 3C 454.3 in the northern hemisphere and 3C 279 in the southern hemisphere look the most promising~\cite{Atoyan:2001ey}.  The recent stacking analysis~\cite{Wang:2015woa} suggest that FSRQs are subdominant as the main sources of the diffuse neutrino flux, which is consistent with our independent conclusion in the previous section.
For BL Lac objects, the IceCube Collaboration has searched for a cross correlation with bright blazars found by {\it Fermi}, and placed an upper limit on the blazar contribution to the diffuse neutrino flux~\cite{Glusenkamp:2015jca}.  As noted above, one possible caveat in such cross-correlation analyses is that weighting each source depends on theoretical modeling especially for distant blazars that are not well studied.  As a complementary check, we also consider one of the brightest BL Lac objects in the northern sky, Mrk 421, and calculate the number of muon events expected in IceCube.  The nondetection of Mrk 421 as well as 3C 273 and 3C 279 as neutrino point sources also supports our conclusion obtained in the previous section (see also Refs.~\cite{Aartsen:2015yva,Aartsen:2016oji}).

The other models listed in Table~1 are unconstrained so far.  But we expect that stacking and cross-correlation analyses are promising for (iii) the SBG model by using catalogues obtained at the infrared or gamma-ray band, (iv) the GC/GG model by using x-ray catalogues, (v) the RL AGN model by using catalogues obtained at the radio or gamma-ray band, (vi) the RQ AGN model by using x-ray catalogues, and (vii) the LL AGN model by using optical or x-ray catalogues.

\begin{figure}[t]
\includegraphics[width=3.00in]{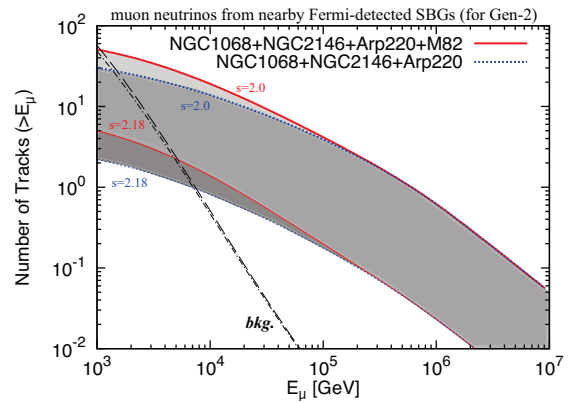}
\caption{The number of muon tracks expected in stacking four representative SBGs, Arp 220, NGC 1068 and 2146, and a prototype SBG, M 82.  Five years of operations by {\it IceCube-Gen2}-like detectors are assumed.  Muon neutrino fluxes are normalized by the observed gamma-ray fluxes at GeV energies~\cite{Ackermann:2012vca,Tang:2014dia}, for $s=2.0$ (thick) and $s=2.18$ (thin), respectively.  Here, the observational ($1\sigma$) uncertainties in the gamma-ray fluxes are indicated by the shaded regions.  The background curves (dashed and dot-dashed) include both the atmospheric and astrophysical neutrino backgrounds.
}
\label{figSBG}
\vspace{-1.\baselineskip}
\end{figure}

\begin{figure}[t]
\includegraphics[width=3.00in]{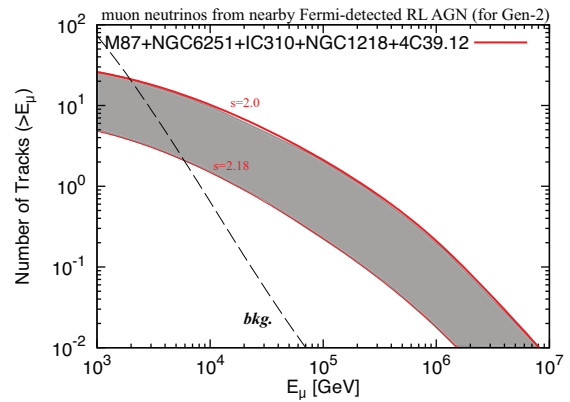}
\caption{The number of muon tracks expected in stacking five nearby RL AGN, M 87, NGC 6251, IC 310, NGC 1218, and 4C +39.12.  As in Fig.~\ref{figSBG}, five years of operations by {\it IceCube-Gen2}-like detectors are assumed.  The background curves (dashed and dot-dashed) include both the atmospheric and astrophysical neutrino backgrounds.  Note that the $pp$ scenario is assumed here, so the detectability of neutrinos is much smaller in the standard leptonic scenario.
}
\label{figRLAGN}
\vspace{-1.\baselineskip}
\end{figure}

In particular, SBGs are intriguing since nearby SBGs are have been detected in gamma rays.
Among nearby SBGs detected by {\it Fermi}, Arp 220 is an ultraluminous infrared galaxy with $L_{\rm IR}\approx1.4\times{10}^{12}L_\odot$, whose gamma rays is recently discovered~\cite{Peng:2016nsx,Griffin:2016wzb}.  Also, NGC 1068 and NGC 2146 with $L_{\rm IR}\approx(1-3)\times{10}^{11}L_\odot$, which is comparable to the infrared luminosity of representative SBGs, $L_{\rm IR}^{\rm eff}\sim3\times{10}^{11}L_\odot$, so we find that they should be regarded as promising neutrino sources in the ``calorimetric'' SBG model.  Using $E_\gamma L_{E_\gamma}\approx2(E_\gamma/2E_\nu)^{2-s}E_\nu L_{E_{\nu_\mu}}$ (expected in $pp$ scenarios; see below), we calculate their neutrino fluxes and evaluate detection rates of muon tracks.  The results are shown in Fig.~\ref{figSBG}, showing that {\it IceCube-Gen2} can detect signals around these nearby SBGs in several years.  In particular, the $s=2.0$ case that safely explains the high-energy IceCube data (see Fig.~\ref{Unified}) is promising.  Note that NGC 1068 coexists with an AGN (Seyfert galaxy), which is important to test the SBG-AGN model suggested by Refs.~\cite{Murase:2014foa,Tamborra:2014xia}.  
The number of $>5$~TeV muon tracks expected in ten years of operations by {\it IceCube-Gen2} is ${\mathcal N}_\mu\sim20$ for $s=2.0$ and ${\mathcal N}_\mu\sim3$ for $s=2.18$, respectively. 
We also consider the possible contribution of M 82.  However, M 82 (in the northern sky) and NGC 253 (in the southern sky) are prototypical SBGs with $L_{\rm IR}\sim{\rm a~few}\times{10}^{10}L_\odot$, which may be so compact that the confinement of $\sim50-100$~PeV protons may be difficult and they may not be ideal PeV neutrino emitters.
Note that NGC 4945 is also detected with gamma rays but located in the southern sky.

Next, we discuss the detectability of RL AGN assuming that neutrinos and gamma rays are produced by $pp$ interactions~\cite{Murase:2013rfa,Tjus:2014dna,Hooper:2016jls} (although the canonical picture for MeV-GeV emission from this class of AGN is the leptonic scenario).  Among the sources listed in the 3FGL catalog, Cen A, Cen B, Pic A, and PKS 0625-35 are located in the southern sky.  On the other hand, ten RL AGN (NGC 1275, NGC 6251, M 87, 3C 111, IC 310, NGC 1218, 4C +39.12, 3C 264, NGC 2484, and 3C 303) are in the northern sky.  NGC 1275 is the brightest in gamma rays, but it is highly variable so that the gamma-ray emission cannot be attributed to the host galaxy or the environment of the Perseus cluster.  Also, the observed gamma-ray spectrum of 3C 111 is too steep for CR reservoir models. In this work, excluding NGC 1275, we consider the five brightest RL AGN with $s\sim2.0-2.2$ in the northern sky, NGC 6251, M 87, IC 310, NGC 1218, and 4C +39.12 that are listed in the 3FGL catalog~\cite{Acero:2015hja}.  The results are shown in Fig.~\ref{figRLAGN}, which imply that {\it IceCube-Gen2} would detect signals around these nearby RL AGN in several years if the $pp$ scenario is correct.  Nondetection of high-energy neutrinos correlated with RL AGN will give us useful constraints complementary to the limits from neutrino multiplet searches, and will support a leptonic origin of gamma-ray emission from RL AGN.

So far we have considered SBGs and RL AGN since they are detected as gamma-ray point sources.  Note that, if GCs/GGs contribute to IceCube's flux, nearby GCs such as the Virgo cluster are detectable as single neutrino sources by next-generation detectors such as {\it IceCube-Gen2}~\cite{Murase:2012rd}.

Whereas the ``statistical'' detection of neutrino sources seems the fastest way, the robust identification of a single source is desirable for the neutrino astronomy.  However, this may be challenging even for {\it IceCube-Gen2}.  The lower limit on the number density of sources implies that the number of neutrino sources contributing to the flux is $\gtrsim10^6$, and that the angular source density is $\gtrsim30~{\rm deg}^{-2}$.  If the angular uncertainty in the determination of the direction of a neutrino-induced muon track is $\sim0.1$~deg, the number of sources located along the line of sight consistent with the neutrino arrival direction is $\gtrsim1(\Delta\theta/0.1{\rm deg})^2$.  This in turn implies that the sources may not be identified by searching for an angular correlation between the neutrino arrival direction and the location on the sky of various astrophysical objects.

\section{Searching for the brightest neutrino sources: gamma rays}\label{multi2}
Since both charged and neutral pions are produced by hadronic interactions of high-energy nucleons with target photons or nucleons, the emission of neutrinos by charged pion decay is accompanied by the emission of gamma rays by neutral pion decay.  The characteristic photon energy is roughly twice the characteristic neutrino energy, and the gamma-ray energy production rate is approximately the same as the neutrino production rate (the exact ratio depends on the particle and radiation spectra).  The neutrino sources are therefore expected to also be gamma-ray sources of similar luminosity.

Assuming that the parent CRs are produced with a power-law spectrum, $dN_{\rm cr}/dE_{\rm cr}\propto E_{\rm cr}^{-s}$, and that the meson production is dominated by inelastic $pp$ collisions with nucleons, the differential gamma-ray luminosity is expected to be $E_\gamma L_{E_\gamma}\approx2(E_\gamma/2E_\nu)^{2-s}E_\nu L_{E_{\nu_\mu}}$~\cite{Murase:2013rfa}.  This is due to the fact that the energy-loss time to meson production is not strongly dependent on the energy of the CRs and the CR spectrum is not too hard, e.g., the confinement time of CRs within an environment in which they may undergo inelastic $pp$ collisions is expected to decrease with energy.  If, on the other hand, meson production is dominated by interactions of CRs with radiation, the gamma-ray luminosity at low energies may be well below that of the high-energy neutrino luminosity.  This is partly due to the fact that the energy threshold for pion production in interactions with radiation fields, $E_p E_\gamma\gtrsim0.2~\Gamma^2~{\rm GeV}^2$ (where $\Gamma$ is the bulk Lorentz factor) for the photomeson production, may be above the energy of the CRs for which the production of pions would lead to gamma rays at observable energies~\cite{Murase:2015xka}.

Note that the gamma-ray luminosity may be suppressed by two effects.
First, high-energy gamma rays may be absorbed by two-photon annihilation interactions with the radiation field in or around the source.
Ref.~\cite{Murase:2015xka} showed that this internal attenuation should naturally occur if IceCube's neutrinos are produced via $p\gamma$ interactions.
In what follows we assume that this effect is negligible, and this assumption is valid for candidate sources like SBGs and GCs/GGs up to $\sim10-100$~TeV energies.  Second, high-energy gamma rays may be absorbed by two-photon annihilation interactions with the extragalactic background light including the cosmic optical and infrared backgrounds.  We take this effect, which suppresses the flux of $\gtrsim0.1-0.3$~TeV gamma rays from cosmologically distant ($d_L\gtrsim1-3$~Gpc) sources, in a manner similar to that of Refs.~\cite{Murase:2012df,Kneiske:2003tx}.

As shown above, CR reservoir models are promising targets for {\it IceCube-Gen2}.  The muon neutrino constraints will reach $n_0^{\rm eff}$ and $n_0^{\rm tot}$ indicated in Table~1 after $\sim5-10$~yr observations.  However, even if the statistical detection is possible, the robust identification of a single neutrino source may be difficult due to the difficulty in excluding many distant candidate sources.  Thus, establishing its gamma-ray counterpart is important to have convincing evidence of a single neutrino source detection.  Firstly, the angular resolution of imaging Cherenkov telescopes is better.  For example, the Cherenkov Telescope Array (CTA)~\cite{Consortium:2010bc} will achieve $\sim0.05$~deg at TeV energies and even smaller at higher energies.  Secondly, multi-TeV sources should be local sources because of the attenuation due to the extragalactic background light.  For example, the High-Altitude Water Cherenkov Observatory (HAWC)~\cite{Abeysekara:2013tza} may detect $\sim10-100$~TeV gamma-ray sources within $\sim100$~Mpc.

\begin{figure}[t]
\includegraphics[width=3.00in]{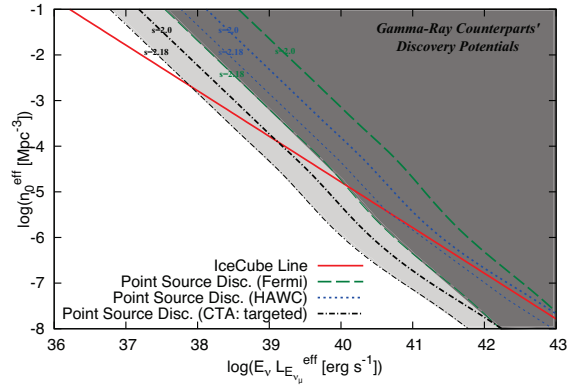}
\caption{The local ($z=0$) number density of neutrino sources, whose gamma-ray counterparts can be discovered by the current {\it Fermi} (with eight-year observation), HAWC (with five-year observation) and future CTA (with 50~hr observation per source).  We consider $pp$ sources with $E_\gamma L_{E_\gamma}\approx2(E_\gamma/2E_\nu)^{2-s}E_\nu L_{E_{\nu_\mu}}$ (see text for details).
The solid red line corresponds to the neutrino luminosity density indicated by the IceCube observation, as indicated by Eq.~(\ref{eq:n0L}). The SFR evolution is assumed.
}
\label{figgam}
\vspace{-1.\baselineskip}
\end{figure}

To see whether gamma-ray counterparts of single neutrino sources (that will be inferred by IceCube or {\it IceCube-Gen2}) can be discovered or not, in Fig.~\ref{figgam}, we show the number density of candidate neutrino sources, whose gamma-ray spectra can be measured by various gamma-ray experiments including the current {\it Fermi}, HAWC, and future CTA.
The $5\sigma$ significance discovery potential for point sources is used.
We consider $2.0\leq s\lesssim2.2$. The upper limit on $s$ is set by the isotropic diffuse gamma-ray background measured in the $0.1-820$~GeV range (gamma-ray sources with larger values of $s$ that explain the observed IceCube neutrino intensity produce a gamma-ray background violating the {\it Fermi} data~\cite{Murase:2013rfa}).  Using Eq.~(\ref{eq:Nm2}) for CR reservoir models, the number density of neutrino sources reachable by gamma-ray detectors is approximately given by
\begin{eqnarray}
n_0^{\rm eff}&\sim&2\times{10}^{-5}~{\rm Mpc}^{-3}~{\left(\frac{E_\gamma^2 \Phi_{\gamma}}{2\times{10}^{-8}~{\rm GeV}~{\rm cm}^{-2}~{\rm s}^{-1}~{\rm sr}^{-1}}\right)}^3\nonumber\\&\times&{\left(\frac{\xi_z}{3}\right)}^{-3}{\left(\frac{F_{\rm lim}}{{10}^{-10}~{\rm GeV}~{\rm cm}^{-2}~{\rm s}^{-1}}\right)}^{-3}{\left(\frac{{\Delta \Omega}}{2\pi}\right)}^{2}.\,\,\,\,\,\,\,\,
\end{eqnarray}
For {\it Fermi-LAT} ($0.1-300$~GeV) and HAWC ($0.3-100$~TeV), which are observatories with a wide field of view, their discovery potentials imply that SBGs and GCs/GGs, predicting $n_0^{\rm eff}\sim{10}^{-5}~{\rm Mpc}^{-3}$, can be discovered for $s\sim2.2$.
Note that {\it Fermi}'s all-sky survey should have yielded a detection of a few sources for sources with $s\sim2.2$ and density of $\sim10^{-5}~{\rm Mpc}^{-3}$, as expected for SBGs and GCs/GGs. Indeed, high-energy gamma-ray emission from several nearby SBGs has been detected~\cite{Ackermann:2012vca,Tang:2014dia}, consistent with the prediction of the SBG model in which SBGs are the sources of IceCube's neutrinos.  The nondetection of GCs/GGs does not yet rule out these objects as candidate sources, since nearby objects of this type are extended (for {\it Fermi}'s resolution), and the flux sensitivity for extended sources is worse than that for point sources.

For CTA ($0.02-300$~TeV), which is a narrow field-of-view observatory, the single source discovery line refers to a study of catalogs of known sources which are suggested as neutrino source candidates, assuming 50~hr integration per source. We do not assume the survey mode.
Fig.~\ref{figgam} implies that, if SBGs or GCs/GGs or perhaps RL AGN are responsible for the observed high-energy neutrino flux, single neutrino source candidates found by {\it IceCube-Gen2} via, e.g., multiplet or stacking analyses should be discovered with multi-TeV gamma-ray observations (even for a hard spectral index $s=2.0$).  We note that follow-up observations of high-energy muon neutrino events would also be useful. 

Among the nearby ($<100$~Mpc) SBGs in the catalogue used in Ref.~\cite{Ackermann:2012vca}, 18 SBGs have $L_{\rm IR}\gtrsim{10}^{11}L_\odot$, which can be representative neutrino sources in the ``calorimetric'' SBG model.  The promising targets in the northern sky include NGC 2146, NGC 1068, Arp 299, NGC 6701, NGC 7771, NGC 7469, Arp 220, Mrk 331, NGC 828, Arp 193, and NGC 6240, which can be detected by CTA if SBGs are the sources of IceCube's neutrinos.

For RL AGN, all 3FGL sources will be promising targets for CTA.  An important test is the measurement of time variability. If neutrinos and gamma rays are produced via inelastic $pp$ interactions in their host galaxies or cluster environment, significant variability is not expected. Variable gamma-ray emission can exclude CR reservoir models for RL AGN, and will favor the emission from core regions (where the internal attenuation may be relevant).

Finally, we note that a lower limit on the source density may be obtained from the upper limit on the anisotropy in the extragalactic gamma-ray background measured by {\it Fermi} ($C_p\leq2\times{10}^{-20}~{\rm cm}^{-4}~{\rm s}^{-2}~{\rm sr}^{-1}$ at 20~GeV~\cite{Cuoco:2012yf}, where $C_p$ is the angular power spectrum).
The recent results obtained via the photon count fluctuation analyses~\cite{TheFermi-LAT:2015ykq,Lisanti:2016jub,Zechlin:2015wdz,Zechlin:2016pme} can be used for additional constraints, and the cross correlation gives stringent limits on contributions from star-forming galaxies including SBGs~\cite{Ando:2015bva}.

\section{Discussion and summary}\label{sec:discussion}
We have derived in Sec.~\ref{nuA} constraints on the density and luminosity of steady ``standard candle" neutrino sources dominating the high-energy, $\gtrsim100$~TeV, neutrino flux detected in IceCube, based on the nondetection of  ``point sources" producing high-energy multiple neutrino-induced muon tracks in the detector.  The limits are given in Eqs.~(\ref{eq:Lnu}) and(\ref{eq:n0}), and illustrated in Figs.~\ref{IClim1} and \ref{IClim2} (an upper limit on the density of steady sources at a given luminosity, which is valid for sources that do not necessarily dominate the flux, is given in Eq.~\ref{eq:n0a}).

These limits were applied in Sec.~\ref{nuB} to a wide range of potential source classes, taking into account their redshift evolution and LF.  While the distribution of electromagnetic luminosities, i.e. the photon LF, of different classes of objects are known, the neutrino LFs of most source classes are not known and are model dependent. We therefore did not attempt a comprehensive analysis under different model assumptions regarding the neutrino LFs of various classes of objects. Rather, for each class of objects we defined an effective number density $n_s^{\rm eff}$ (see Eq.~\ref{eq:neff}), characterizing the density of sources dominating the flux.  Our conclusions are not sensitive to the details of the relation between the photon and the neutrino LFs and to the exact definition of $n_s^{\rm eff}$. The classes of sources that are ruled out by the constraint of Eq.~(\ref{eq:n0}), and for which there is a large difference between $n_0^{\rm eff}$ and the total number density $n_0^{\rm tot}$ (see Table 1), are characterized by $n_0^{\rm eff}$ values which are orders of magnitude smaller than the limit of Eq.~(\ref{eq:n0}).

The constraints imply that rare objects, such as powerful blazar jets, are unlikely to dominate IceCube's flux.  For blazars, we showed that the conclusion does not change even if harder (possibly more realistic) neutrino spectra are used (see Fig.~\ref{IClim2}).  This result is consistent with those obtained from stacking and cross-correlation analyses~\cite{Wang:2015woa,Glusenkamp:2015jca}.
However, it should be noted that neutrino emission by transient AGN ``flares" \cite{Atoyan:2001ey,Dermer:2014vaa,Petropoulou:2016ujj,Kadler:2016ygj} is not constrained by the current analysis, as is the case for other types of transient sources.

CR reservoir models and AGN core models, with source density of $n_0^{\rm eff}\gtrsim3\times10^{-6}~{\rm Mpc}^{-3}$, are not constrained by current IceCube data.  An order of magnitude improvement in $F_{\rm lim}$, the minimum flux required for a source to be detectable as a point source, can improve the limit on $n_0$ by more than two orders of magnitude, and will likely enable the detection (as point sources) of the few brightest objects for almost all candidate source classes, including SBGs, RL AGN, and GCs/GGs (see Table 1 and Fig.~\ref{IClim1}).  Such an improvement in $F_{\rm lim}$ requires an order of magnitude increase in the effective mass of the detector at $0.1-1$~PeV (where the background is negligible), which may be provided by {\it IceCube-Gen2} and an upgraded KM3NeT.

Searches for the brightest neutrino sources, including stacking and cross-correlation analyses, are powerful especially for the SBG and RL AGN models.  However, in general, they are model dependent.  While the detection of a few point sources may confirm the validity of a suggested source model, nondetections may not necessarily rule out all the models for the suggested source class.  This is due to the fact that large deviations from an ``average source luminosity" cannot be excluded when the source physics is not well understood, and model uncertainties often prevent accurate predictions. For example, testing the LL AGN core model is feasible in the canonical case since $n_0^{\rm eff}\sim{10}^{-3}~{\rm Mpc}^{-3}$ can be reached by {\it IceCube-Gen2} for nonevolving sources.  However, accessing $n_0^{\rm tot}\gtrsim{10}^{-2}~{\rm Mpc}^{-3}$ may be difficult.

At photon energies of 1~GeV to 1~TeV, which are well below the energy of the neutrinos observed by IceCube but accessible to gamma-ray telescopes, a gamma-ray luminosity of $E_\gamma L_{E_\gamma}\approx2(E_\gamma/2E_\nu)^{2-s}E_\nu L_{E_{\nu_\mu}}$ is expected for CR reservoirs (like SBGs and GCs/GGs) in which (a) the parent CRs are produced with a power-law spectrum, (b) the production of mesons is dominated by inelastic $pp$ collisions with nucleons, and (c) the internal absorption of gamma rays by two-photon annihilation interactions is negligible below $\sim1-10$~TeV.
We showed that gamma-ray observations may be useful for testing models of this type. In particular, dedicated targeted observations by the CTA detector of the brightest objects of a complete catalogue of candidate neutrino sources will lead to the detection of individual bright sources for source classes with $n_0^{\rm eff}\lesssim10^{-4}~{\rm Mpc}^{-3}$.


\medskip
\begin{acknowledgments}
K.~M. acknowledges the hospitality of Institute for Advanced Study and Weizmann Institute of Science.  K.~M. thanks Markus Ahlers, Shin'ichiro Ando, Kfir Blum, Doug Cowen, Chuck Dermer, Francis Halzen, Yoshiyuki Inoue, Shigeo Kimura, Kumiko Kotera, Brian Lacki, Peter M\'esz\'aros, Naoko Kurahashi Neilson, Foteini Oikonomou, Anna Stasto, Shigeru Yoshida, and Fabio Zandanel for useful discussion.
The work of K.~M. is supported by NSF Grant No. PHY-1620777.  E.~W. is partially supported by BSF and ISF-I-Core grants.
We first presented our results in the JSI workshop on Multimessenger Astronomy in the Era of PeV Neutrinos in November 2014~\cite{Murase:2014JSI}.
\end{acknowledgments}


\bibliography{kmurase}

\hyphenation{Post-Script Sprin-ger}
\begin{thebibliography}{122}
\expandafter\ifx\csname natexlab\endcsname\relax\def\natexlab#1{#1}\fi
\expandafter\ifx\csname bibnamefont\endcsname\relax
  \def\bibnamefont#1{#1}\fi
\expandafter\ifx\csname bibfnamefont\endcsname\relax
  \def\bibfnamefont#1{#1}\fi
\expandafter\ifx\csname citenamefont\endcsname\relax
  \def\citenamefont#1{#1}\fi
\expandafter\ifx\csname url\endcsname\relax
  \def\url#1{\texttt{#1}}\fi
\expandafter\ifx\csname urlprefix\endcsname\relax\def\urlprefix{URL }\fi
\providecommand{\bibinfo}[2]{#2}
\providecommand{\eprint}[2][]{\url{#2}}

\bibitem[{\citenamefont{Aartsen et~al.}(2013{\natexlab{a}})}]{Aartsen:2013bka}
\bibinfo{author}{\bibfnamefont{M.}~\bibnamefont{Aartsen}} \bibnamefont{et~al.}
  (\bibinfo{collaboration}{IceCube Collaboration}),
  \bibinfo{journal}{Phys.Rev.Lett.} \textbf{\bibinfo{volume}{111}},
  \bibinfo{pages}{021103} (\bibinfo{year}{2013}{\natexlab{a}}),
  \eprint{1304.5356}.

\bibitem[{\citenamefont{Aartsen et~al.}(2013{\natexlab{b}})}]{Aartsen:2013jdh}
\bibinfo{author}{\bibfnamefont{M.}~\bibnamefont{Aartsen}} \bibnamefont{et~al.}
  (\bibinfo{collaboration}{IceCube Collaboration}), \bibinfo{journal}{Science}
  \textbf{\bibinfo{volume}{342}}, \bibinfo{pages}{1242856}
  (\bibinfo{year}{2013}{\natexlab{b}}), \eprint{1311.5238}.

\bibitem[{\citenamefont{Aartsen et~al.}(2014{\natexlab{a}})}]{Aartsen:2014gkd}
\bibinfo{author}{\bibfnamefont{M.}~\bibnamefont{Aartsen}} \bibnamefont{et~al.}
  (\bibinfo{collaboration}{IceCube Collaboration}),
  \bibinfo{journal}{Phys.Rev.Lett.} \textbf{\bibinfo{volume}{113}},
  \bibinfo{pages}{101101} (\bibinfo{year}{2014}{\natexlab{a}}),
  \eprint{1405.5303; https://icecube.wisc.edu/science/data/HE-nu-2010-2014}.

\bibitem[{\citenamefont{Aartsen et~al.}(2015{\natexlab{a}})}]{Aartsen:2014muf}
\bibinfo{author}{\bibfnamefont{M.}~\bibnamefont{Aartsen}} \bibnamefont{et~al.}
  (\bibinfo{collaboration}{IceCube Collaboration}),
  \bibinfo{journal}{Phys.Rev.} \textbf{\bibinfo{volume}{D91}},
  \bibinfo{pages}{022001} (\bibinfo{year}{2015}{\natexlab{a}}),
  \eprint{1410.1749}.

\bibitem[{\citenamefont{Aartsen et~al.}(2015{\natexlab{b}})}]{Aartsen:2015ita}
\bibinfo{author}{\bibfnamefont{M.~G.} \bibnamefont{Aartsen}}
  \bibnamefont{et~al.} (\bibinfo{collaboration}{IceCube Collaboration}),
  \bibinfo{journal}{Astrophys. J.} \textbf{\bibinfo{volume}{809}},
  \bibinfo{pages}{98} (\bibinfo{year}{2015}{\natexlab{b}}),
  \eprint{1507.03991}.

\bibitem[{\citenamefont{Aartsen et~al.}(2015{\natexlab{c}})}]{Aartsen:2015rwa}
\bibinfo{author}{\bibfnamefont{M.~G.} \bibnamefont{Aartsen}}
  \bibnamefont{et~al.} (\bibinfo{collaboration}{IceCube Collaboration}),
  \bibinfo{journal}{Phys. Rev. Lett.} \textbf{\bibinfo{volume}{115}},
  \bibinfo{pages}{081102} (\bibinfo{year}{2015}{\natexlab{c}}),
  \eprint{1507.04005}.

\bibitem[{\citenamefont{Waxman and Bahcall}(1998)}]{Waxman:1998yy}
\bibinfo{author}{\bibfnamefont{E.}~\bibnamefont{Waxman}} \bibnamefont{and}
  \bibinfo{author}{\bibfnamefont{J.~N.} \bibnamefont{Bahcall}},
  \bibinfo{journal}{Phys.Rev.} \textbf{\bibinfo{volume}{D59}},
  \bibinfo{pages}{023002} (\bibinfo{year}{1998}), \eprint{hep-ph/9807282}.

\bibitem[{\citenamefont{Waxman}(2013)}]{Waxman:2013zda}
\bibinfo{author}{\bibfnamefont{E.}~\bibnamefont{Waxman}}
  (\bibinfo{year}{2013}), \eprint{1312.0558}.

\bibitem[{\citenamefont{Halzen}(2014)}]{Halzen:2013dva}
\bibinfo{author}{\bibfnamefont{F.}~\bibnamefont{Halzen}},
  \bibinfo{journal}{Nuovo Cim.} \textbf{\bibinfo{volume}{C037}},
  \bibinfo{pages}{117} (\bibinfo{year}{2014}), \bibinfo{note}{[Astron.
  Nachr.335,507(2014)]}, \eprint{1311.6350}.

\bibitem[{\citenamefont{M\'esz\'aros}(2014)}]{Meszaros:2014tta}
\bibinfo{author}{\bibfnamefont{P.}~\bibnamefont{M\'esz\'aros}},
  \bibinfo{journal}{Nucl. Phys. Proc. Suppl.}
  \textbf{\bibinfo{volume}{256-257}}, \bibinfo{pages}{241}
  (\bibinfo{year}{2014}), \eprint{1407.5671}.

\bibitem[{\citenamefont{Murase and Beacom}(2010)}]{Murase:2010gj}
\bibinfo{author}{\bibfnamefont{K.}~\bibnamefont{Murase}} \bibnamefont{and}
  \bibinfo{author}{\bibfnamefont{J.~F.} \bibnamefont{Beacom}},
  \bibinfo{journal}{Phys.Rev.} \textbf{\bibinfo{volume}{D81}},
  \bibinfo{pages}{123001} (\bibinfo{year}{2010}), \eprint{1003.4959}.

\bibitem[{\citenamefont{Aartsen et~al.}(2015{\natexlab{d}})}]{Aartsen:2015zva}
\bibinfo{author}{\bibfnamefont{M.~G.} \bibnamefont{Aartsen}}
  \bibnamefont{et~al.} (\bibinfo{collaboration}{IceCube Collaboration}), in
  \emph{\bibinfo{booktitle}{{Proceedings, 34th International Cosmic Ray
  Conference (ICRC 2015)}}} (\bibinfo{year}{2015}{\natexlab{d}}),
  \eprint{1510.05223}.

\bibitem[{\citenamefont{Murase et~al.}(2016)\citenamefont{Murase, Guetta, and
  Ahlers}}]{Murase:2015xka}
\bibinfo{author}{\bibfnamefont{K.}~\bibnamefont{Murase}},
  \bibinfo{author}{\bibfnamefont{D.}~\bibnamefont{Guetta}}, \bibnamefont{and}
  \bibinfo{author}{\bibfnamefont{M.}~\bibnamefont{Ahlers}},
  \bibinfo{journal}{Phys. Rev. Lett.} \textbf{\bibinfo{volume}{116}},
  \bibinfo{pages}{071101} (\bibinfo{year}{2016}), \eprint{1509.00805}.

\bibitem[{\citenamefont{Chen et~al.}(2015)\citenamefont{Chen, Bhupal~Dev, and
  Soni}}]{Chen:2014gxa}
\bibinfo{author}{\bibfnamefont{C.-Y.} \bibnamefont{Chen}},
  \bibinfo{author}{\bibfnamefont{P.~S.} \bibnamefont{Bhupal~Dev}},
  \bibnamefont{and} \bibinfo{author}{\bibfnamefont{A.}~\bibnamefont{Soni}},
  \bibinfo{journal}{Phys. Rev.} \textbf{\bibinfo{volume}{D92}},
  \bibinfo{pages}{073001} (\bibinfo{year}{2015}), \eprint{1411.5658}.

\bibitem[{\citenamefont{Palladino and Vissani}(2016)}]{Palladino:2016zoe}
\bibinfo{author}{\bibfnamefont{A.}~\bibnamefont{Palladino}} \bibnamefont{and}
  \bibinfo{author}{\bibfnamefont{F.}~\bibnamefont{Vissani}},
  \bibinfo{journal}{Astrophys. J.} \textbf{\bibinfo{volume}{826}},
  \bibinfo{pages}{185} (\bibinfo{year}{2016}), \eprint{1601.06678}.

\bibitem[{\citenamefont{Neronov and Semikoz}(2016)}]{Neronov:2016bnp}
\bibinfo{author}{\bibfnamefont{A.}~\bibnamefont{Neronov}} \bibnamefont{and}
  \bibinfo{author}{\bibfnamefont{D.}~\bibnamefont{Semikoz}},
  \bibinfo{journal}{Phys. Rev.} \textbf{\bibinfo{volume}{D93}},
  \bibinfo{pages}{123002} (\bibinfo{year}{2016}), \eprint{1603.06733}.

\bibitem[{\citenamefont{Ackermann et~al.}(2015)}]{Ackermann:2014usa}
\bibinfo{author}{\bibfnamefont{M.}~\bibnamefont{Ackermann}}
  \bibnamefont{et~al.} (\bibinfo{collaboration}{Fermi LAT collaboration}),
  \bibinfo{journal}{Astrophys.J.} \textbf{\bibinfo{volume}{799}},
  \bibinfo{pages}{86} (\bibinfo{year}{2015}), \eprint{1410.3696}.

\bibitem[{\citenamefont{Murase et~al.}(2013)\citenamefont{Murase, Ahlers, and
  Lacki}}]{Murase:2013rfa}
\bibinfo{author}{\bibfnamefont{K.}~\bibnamefont{Murase}},
  \bibinfo{author}{\bibfnamefont{M.}~\bibnamefont{Ahlers}}, \bibnamefont{and}
  \bibinfo{author}{\bibfnamefont{B.~C.} \bibnamefont{Lacki}},
  \bibinfo{journal}{Phys.Rev.} \textbf{\bibinfo{volume}{D88}},
  \bibinfo{pages}{121301} (\bibinfo{year}{2013}), \eprint{1306.3417}.

\bibitem[{\citenamefont{Loeb and Waxman}(2006)}]{Loeb:2006tw}
\bibinfo{author}{\bibfnamefont{A.}~\bibnamefont{Loeb}} \bibnamefont{and}
  \bibinfo{author}{\bibfnamefont{E.}~\bibnamefont{Waxman}},
  \bibinfo{journal}{JCAP} \textbf{\bibinfo{volume}{0605}}, \bibinfo{pages}{003}
  (\bibinfo{year}{2006}), \eprint{astro-ph/0601695}.

\bibitem[{\citenamefont{Katz et~al.}(2013)\citenamefont{Katz, Waxman, Thompson,
  and Loeb}}]{Katz:2013ooa}
\bibinfo{author}{\bibfnamefont{B.}~\bibnamefont{Katz}},
  \bibinfo{author}{\bibfnamefont{E.}~\bibnamefont{Waxman}},
  \bibinfo{author}{\bibfnamefont{T.}~\bibnamefont{Thompson}}, \bibnamefont{and}
  \bibinfo{author}{\bibfnamefont{A.}~\bibnamefont{Loeb}}
  (\bibinfo{year}{2013}), \eprint{1311.0287}.

\bibitem[{\citenamefont{Waxman and Bahcall}(1997)}]{Waxman:1997ti}
\bibinfo{author}{\bibfnamefont{E.}~\bibnamefont{Waxman}} \bibnamefont{and}
  \bibinfo{author}{\bibfnamefont{J.~N.} \bibnamefont{Bahcall}},
  \bibinfo{journal}{Phys.Rev.Lett.} \textbf{\bibinfo{volume}{78}},
  \bibinfo{pages}{2292} (\bibinfo{year}{1997}), \eprint{astro-ph/9701231}.

\bibitem[{\citenamefont{Murase et~al.}(2006)\citenamefont{Murase, Ioka,
  Nagataki, and Nakamura}}]{Murase:2006mm}
\bibinfo{author}{\bibfnamefont{K.}~\bibnamefont{Murase}},
  \bibinfo{author}{\bibfnamefont{K.}~\bibnamefont{Ioka}},
  \bibinfo{author}{\bibfnamefont{S.}~\bibnamefont{Nagataki}}, \bibnamefont{and}
  \bibinfo{author}{\bibfnamefont{T.}~\bibnamefont{Nakamura}},
  \bibinfo{journal}{Astrophys.J.} \textbf{\bibinfo{volume}{651}},
  \bibinfo{pages}{L5} (\bibinfo{year}{2006}), \eprint{astro-ph/0607104}.

\bibitem[{\citenamefont{Bustamante et~al.}(2015)\citenamefont{Bustamante,
  Baerwald, Murase, and Winter}}]{Bustamante:2014oka}
\bibinfo{author}{\bibfnamefont{M.}~\bibnamefont{Bustamante}},
  \bibinfo{author}{\bibfnamefont{P.}~\bibnamefont{Baerwald}},
  \bibinfo{author}{\bibfnamefont{K.}~\bibnamefont{Murase}}, \bibnamefont{and}
  \bibinfo{author}{\bibfnamefont{W.}~\bibnamefont{Winter}},
  \bibinfo{journal}{Nature Communications} \textbf{\bibinfo{volume}{6}},
  \bibinfo{pages}{6783} (\bibinfo{year}{2015}), \eprint{1409.2874}.

\bibitem[{\citenamefont{Mannheim}(1995)}]{Mannheim:1995mm}
\bibinfo{author}{\bibfnamefont{K.}~\bibnamefont{Mannheim}},
  \bibinfo{journal}{Astropart.Phys.} \textbf{\bibinfo{volume}{3}},
  \bibinfo{pages}{295} (\bibinfo{year}{1995}).

\bibitem[{\citenamefont{Atoyan and Dermer}(2001)}]{Atoyan:2001ey}
\bibinfo{author}{\bibfnamefont{A.}~\bibnamefont{Atoyan}} \bibnamefont{and}
  \bibinfo{author}{\bibfnamefont{C.~D.} \bibnamefont{Dermer}},
  \bibinfo{journal}{Phys.Rev.Lett.} \textbf{\bibinfo{volume}{87}},
  \bibinfo{pages}{221102} (\bibinfo{year}{2001}), \eprint{astro-ph/0108053}.

\bibitem[{\citenamefont{Atoyan and Dermer}(2003)}]{Atoyan:2002gu}
\bibinfo{author}{\bibfnamefont{A.~M.} \bibnamefont{Atoyan}} \bibnamefont{and}
  \bibinfo{author}{\bibfnamefont{C.~D.} \bibnamefont{Dermer}},
  \bibinfo{journal}{Astrophys.J.} \textbf{\bibinfo{volume}{586}},
  \bibinfo{pages}{79} (\bibinfo{year}{2003}), \eprint{astro-ph/0209231}.

\bibitem[{\citenamefont{Dermer et~al.}(2012)\citenamefont{Dermer, Murase, and
  Takami}}]{Dermer:2012rg}
\bibinfo{author}{\bibfnamefont{C.~D.} \bibnamefont{Dermer}},
  \bibinfo{author}{\bibfnamefont{K.}~\bibnamefont{Murase}}, \bibnamefont{and}
  \bibinfo{author}{\bibfnamefont{H.}~\bibnamefont{Takami}},
  \bibinfo{journal}{Astrophys. J.} \textbf{\bibinfo{volume}{755}},
  \bibinfo{pages}{147} (\bibinfo{year}{2012}), \eprint{1203.6544}.

\bibitem[{\citenamefont{Murase et~al.}(2008)\citenamefont{Murase, Inoue, and
  Nagataki}}]{Murase:2008yt}
\bibinfo{author}{\bibfnamefont{K.}~\bibnamefont{Murase}},
  \bibinfo{author}{\bibfnamefont{S.}~\bibnamefont{Inoue}}, \bibnamefont{and}
  \bibinfo{author}{\bibfnamefont{S.}~\bibnamefont{Nagataki}},
  \bibinfo{journal}{Astrophys.J.} \textbf{\bibinfo{volume}{689}},
  \bibinfo{pages}{L105} (\bibinfo{year}{2008}), \eprint{0805.0104}.

\bibitem[{\citenamefont{Kotera et~al.}(2009)\citenamefont{Kotera, Allard,
  Murase, Aoi, Dubois et~al.}}]{Kotera:2009ms}
\bibinfo{author}{\bibfnamefont{K.}~\bibnamefont{Kotera}},
  \bibinfo{author}{\bibfnamefont{D.}~\bibnamefont{Allard}},
  \bibinfo{author}{\bibfnamefont{K.}~\bibnamefont{Murase}},
  \bibinfo{author}{\bibfnamefont{J.}~\bibnamefont{Aoi}},
  \bibinfo{author}{\bibfnamefont{Y.}~\bibnamefont{Dubois}},
  \bibnamefont{et~al.}, \bibinfo{journal}{Astrophys.J.}
  \textbf{\bibinfo{volume}{707}}, \bibinfo{pages}{370} (\bibinfo{year}{2009}),
  \eprint{0907.2433}.

\bibitem[{\citenamefont{Kimura et~al.}(2015)\citenamefont{Kimura, Murase, and
  Toma}}]{Kimura:2014jba}
\bibinfo{author}{\bibfnamefont{S.~S.} \bibnamefont{Kimura}},
  \bibinfo{author}{\bibfnamefont{K.}~\bibnamefont{Murase}}, \bibnamefont{and}
  \bibinfo{author}{\bibfnamefont{K.}~\bibnamefont{Toma}},
  \bibinfo{journal}{Astrophys.J.} \textbf{\bibinfo{volume}{806}},
  \bibinfo{pages}{159} (\bibinfo{year}{2015}), \eprint{1411.3588}.

\bibitem[{\citenamefont{Hooper}(2016)}]{Hooper:2016jls}
\bibinfo{author}{\bibfnamefont{D.}~\bibnamefont{Hooper}},
  \bibinfo{journal}{JCAP} \textbf{\bibinfo{volume}{1609}}, \bibinfo{pages}{002}
  (\bibinfo{year}{2016}), \eprint{1605.06504}.

\bibitem[{\citenamefont{Becker~Tjus et~al.}(2014)\citenamefont{Becker~Tjus,
  Eichmann, Halzen, Kheirandish, and Saba}}]{Tjus:2014dna}
\bibinfo{author}{\bibfnamefont{J.}~\bibnamefont{Becker~Tjus}},
  \bibinfo{author}{\bibfnamefont{B.}~\bibnamefont{Eichmann}},
  \bibinfo{author}{\bibfnamefont{F.}~\bibnamefont{Halzen}},
  \bibinfo{author}{\bibfnamefont{A.}~\bibnamefont{Kheirandish}},
  \bibnamefont{and} \bibinfo{author}{\bibfnamefont{S.~M.} \bibnamefont{Saba}},
  \bibinfo{journal}{Phys. Rev.} \textbf{\bibinfo{volume}{D89}},
  \bibinfo{pages}{123005} (\bibinfo{year}{2014}), \eprint{1406.0506}.

\bibitem[{\citenamefont{Murase et~al.}(2014)\citenamefont{Murase, Inoue, and
  Dermer}}]{Murase:2014foa}
\bibinfo{author}{\bibfnamefont{K.}~\bibnamefont{Murase}},
  \bibinfo{author}{\bibfnamefont{Y.}~\bibnamefont{Inoue}}, \bibnamefont{and}
  \bibinfo{author}{\bibfnamefont{C.~D.} \bibnamefont{Dermer}},
  \bibinfo{journal}{Phys.Rev.} \textbf{\bibinfo{volume}{D90}},
  \bibinfo{pages}{023007} (\bibinfo{year}{2014}), \eprint{1403.4089}.

\bibitem[{\citenamefont{Padovani et~al.}(2015)\citenamefont{Padovani,
  Petropoulou, Giommi, and Resconi}}]{Padovani:2015mba}
\bibinfo{author}{\bibfnamefont{P.}~\bibnamefont{Padovani}},
  \bibinfo{author}{\bibfnamefont{M.}~\bibnamefont{Petropoulou}},
  \bibinfo{author}{\bibfnamefont{P.}~\bibnamefont{Giommi}}, \bibnamefont{and}
  \bibinfo{author}{\bibfnamefont{E.}~\bibnamefont{Resconi}},
  \bibinfo{journal}{Mon.Not.Roy.Astron.Soc.} \textbf{\bibinfo{volume}{452}},
  \bibinfo{pages}{1877} (\bibinfo{year}{2015}), \eprint{1506.09135}.

\bibitem[{\citenamefont{Murase}(2015)}]{Murase:2015ndr}
\bibinfo{author}{\bibfnamefont{K.}~\bibnamefont{Murase}}
  (\bibinfo{year}{2015}), \eprint{1511.01590}.

\bibitem[{\citenamefont{Dermer et~al.}(2014)\citenamefont{Dermer, Murase, and
  Inoue}}]{Dermer:2014vaa}
\bibinfo{author}{\bibfnamefont{C.~D.} \bibnamefont{Dermer}},
  \bibinfo{author}{\bibfnamefont{K.}~\bibnamefont{Murase}}, \bibnamefont{and}
  \bibinfo{author}{\bibfnamefont{Y.}~\bibnamefont{Inoue}},
  \bibinfo{journal}{JHEAp} \textbf{\bibinfo{volume}{3-4}}, \bibinfo{pages}{29}
  (\bibinfo{year}{2014}), \eprint{1406.2633}.

\bibitem[{\citenamefont{Tavecchio et~al.}(2014)\citenamefont{Tavecchio,
  Ghisellini, and Guetta}}]{Tavecchio:2014xha}
\bibinfo{author}{\bibfnamefont{F.}~\bibnamefont{Tavecchio}},
  \bibinfo{author}{\bibfnamefont{G.}~\bibnamefont{Ghisellini}},
  \bibnamefont{and} \bibinfo{author}{\bibfnamefont{D.}~\bibnamefont{Guetta}},
  \bibinfo{journal}{Astrophys.J.} \textbf{\bibinfo{volume}{793}},
  \bibinfo{pages}{L18} (\bibinfo{year}{2014}).

\bibitem[{\citenamefont{Tavecchio and Ghisellini}(2015)}]{Tavecchio:2014eia}
\bibinfo{author}{\bibfnamefont{F.}~\bibnamefont{Tavecchio}} \bibnamefont{and}
  \bibinfo{author}{\bibfnamefont{G.}~\bibnamefont{Ghisellini}},
  \bibinfo{journal}{Mon.Not.Roy.Astron.Soc.} \textbf{\bibinfo{volume}{451}},
  \bibinfo{pages}{1502} (\bibinfo{year}{2015}), \eprint{1411.2783}.

\bibitem[{\citenamefont{Petropoulou et~al.}(2015)\citenamefont{Petropoulou,
  Dimitrakoudis, Padovani, Mastichiadis, and Resconi}}]{Petropoulou:2015upa}
\bibinfo{author}{\bibfnamefont{M.}~\bibnamefont{Petropoulou}},
  \bibinfo{author}{\bibfnamefont{S.}~\bibnamefont{Dimitrakoudis}},
  \bibinfo{author}{\bibfnamefont{P.}~\bibnamefont{Padovani}},
  \bibinfo{author}{\bibfnamefont{A.}~\bibnamefont{Mastichiadis}},
  \bibnamefont{and} \bibinfo{author}{\bibfnamefont{E.}~\bibnamefont{Resconi}},
  \bibinfo{journal}{Mon. Not. Roy. Astron. Soc.}
  \textbf{\bibinfo{volume}{448}}, \bibinfo{pages}{2412} (\bibinfo{year}{2015}),
  \eprint{1501.07115}.

\bibitem[{\citenamefont{Winter}(2013)}]{Winter:2013cla}
\bibinfo{author}{\bibfnamefont{W.}~\bibnamefont{Winter}},
  \bibinfo{journal}{Phys.Rev.} \textbf{\bibinfo{volume}{D88}},
  \bibinfo{pages}{083007} (\bibinfo{year}{2013}), \eprint{1307.2793}.

\bibitem[{\citenamefont{Kistler et~al.}(2014)\citenamefont{Kistler, Stanev, and
  Yuksel}}]{Kistler:2013my}
\bibinfo{author}{\bibfnamefont{M.~D.} \bibnamefont{Kistler}},
  \bibinfo{author}{\bibfnamefont{T.}~\bibnamefont{Stanev}}, \bibnamefont{and}
  \bibinfo{author}{\bibfnamefont{H.}~\bibnamefont{Yuksel}},
  \bibinfo{journal}{Phys.Rev.} \textbf{\bibinfo{volume}{D90}},
  \bibinfo{pages}{123006} (\bibinfo{year}{2014}), \eprint{1301.1703}.

\bibitem[{\citenamefont{Stecker et~al.}(1991)\citenamefont{Stecker, Done,
  Salamon, and Sommers}}]{Stecker:1991vm}
\bibinfo{author}{\bibfnamefont{F.~W.} \bibnamefont{Stecker}},
  \bibinfo{author}{\bibfnamefont{C.}~\bibnamefont{Done}},
  \bibinfo{author}{\bibfnamefont{M.~H.} \bibnamefont{Salamon}},
  \bibnamefont{and} \bibinfo{author}{\bibfnamefont{P.}~\bibnamefont{Sommers}},
  \bibinfo{journal}{Phys.Rev.Lett.} \textbf{\bibinfo{volume}{66}},
  \bibinfo{pages}{2697} (\bibinfo{year}{1991}).

\bibitem[{\citenamefont{Stecker}(2013)}]{Stecker:2013fxa}
\bibinfo{author}{\bibfnamefont{F.~W.} \bibnamefont{Stecker}},
  \bibinfo{journal}{Phys.Rev.} \textbf{\bibinfo{volume}{D88}},
  \bibinfo{pages}{047301} (\bibinfo{year}{2013}), \eprint{1305.7404}.

\bibitem[{\citenamefont{Kalashev et~al.}(2015)\citenamefont{Kalashev, Semikoz,
  and Tkachev}}]{Kalashev:2015cma}
\bibinfo{author}{\bibfnamefont{O.}~\bibnamefont{Kalashev}},
  \bibinfo{author}{\bibfnamefont{D.}~\bibnamefont{Semikoz}}, \bibnamefont{and}
  \bibinfo{author}{\bibfnamefont{I.}~\bibnamefont{Tkachev}},
  \bibinfo{journal}{J.Exp.Theor.Phys.} \textbf{\bibinfo{volume}{120}},
  \bibinfo{pages}{541} (\bibinfo{year}{2015}).

\bibitem[{\citenamefont{Murase and Ioka}(2013)}]{Murase:2013ffa}
\bibinfo{author}{\bibfnamefont{K.}~\bibnamefont{Murase}} \bibnamefont{and}
  \bibinfo{author}{\bibfnamefont{K.}~\bibnamefont{Ioka}},
  \bibinfo{journal}{Phys.Rev.Lett.} \textbf{\bibinfo{volume}{111}},
  \bibinfo{pages}{121102} (\bibinfo{year}{2013}), \eprint{1306.2274}.

\bibitem[{\citenamefont{Nakar}(2015)}]{Nakar:2015tma}
\bibinfo{author}{\bibfnamefont{E.}~\bibnamefont{Nakar}},
  \bibinfo{journal}{Astrophys. J.} \textbf{\bibinfo{volume}{807}},
  \bibinfo{pages}{172} (\bibinfo{year}{2015}), \eprint{1503.00441}.

\bibitem[{\citenamefont{Senno et~al.}(2016)\citenamefont{Senno, Murase, and
  M\'esz\'aros}}]{Senno:2015tsn}
\bibinfo{author}{\bibfnamefont{N.}~\bibnamefont{Senno}},
  \bibinfo{author}{\bibfnamefont{K.}~\bibnamefont{Murase}}, \bibnamefont{and}
  \bibinfo{author}{\bibfnamefont{P.}~\bibnamefont{M\'esz\'aros}},
  \bibinfo{journal}{Phys. Rev.} \textbf{\bibinfo{volume}{D93}},
  \bibinfo{pages}{083003} (\bibinfo{year}{2016}), \eprint{1512.08513}.

\bibitem[{\citenamefont{Tamborra and Ando}(2016)}]{Tamborra:2015fzv}
\bibinfo{author}{\bibfnamefont{I.}~\bibnamefont{Tamborra}} \bibnamefont{and}
  \bibinfo{author}{\bibfnamefont{S.}~\bibnamefont{Ando}},
  \bibinfo{journal}{Phys. Rev.} \textbf{\bibinfo{volume}{D93}},
  \bibinfo{pages}{053010} (\bibinfo{year}{2016}), \eprint{1512.01559}.

\bibitem[{\citenamefont{M\'esz\'aros and Waxman}(2001)}]{Meszaros:2001ms}
\bibinfo{author}{\bibfnamefont{P.}~\bibnamefont{M\'esz\'aros}}
  \bibnamefont{and} \bibinfo{author}{\bibfnamefont{E.}~\bibnamefont{Waxman}},
  \bibinfo{journal}{Phys.Rev.Lett.} \textbf{\bibinfo{volume}{87}},
  \bibinfo{pages}{171102} (\bibinfo{year}{2001}), \eprint{astro-ph/0103275}.

\bibitem[{\citenamefont{Lipari}(2008)}]{Lipari:2008zf}
\bibinfo{author}{\bibfnamefont{P.}~\bibnamefont{Lipari}},
  \bibinfo{journal}{Phys. Rev.} \textbf{\bibinfo{volume}{D78}},
  \bibinfo{pages}{083011} (\bibinfo{year}{2008}), \eprint{0808.0344}.

\bibitem[{\citenamefont{Silvestri and Barwick}(2010)}]{Silvestri:2009xb}
\bibinfo{author}{\bibfnamefont{A.}~\bibnamefont{Silvestri}} \bibnamefont{and}
  \bibinfo{author}{\bibfnamefont{S.~W.} \bibnamefont{Barwick}},
  \bibinfo{journal}{Phys.Rev.} \textbf{\bibinfo{volume}{D81}},
  \bibinfo{pages}{023001} (\bibinfo{year}{2010}), \eprint{0908.4266}.

\bibitem[{\citenamefont{Murase et~al.}(2012)\citenamefont{Murase, Beacom, and
  Takami}}]{Murase:2012df}
\bibinfo{author}{\bibfnamefont{K.}~\bibnamefont{Murase}},
  \bibinfo{author}{\bibfnamefont{J.~F.} \bibnamefont{Beacom}},
  \bibnamefont{and} \bibinfo{author}{\bibfnamefont{H.}~\bibnamefont{Takami}},
  \bibinfo{journal}{JCAP} \textbf{\bibinfo{volume}{1208}}, \bibinfo{pages}{030}
  (\bibinfo{year}{2012}), \eprint{1205.5755}.

\bibitem[{\citenamefont{Ahlers and Halzen}(2014)}]{Ahlers:2014ioa}
\bibinfo{author}{\bibfnamefont{M.}~\bibnamefont{Ahlers}} \bibnamefont{and}
  \bibinfo{author}{\bibfnamefont{F.}~\bibnamefont{Halzen}},
  \bibinfo{journal}{Phys.Rev.} \textbf{\bibinfo{volume}{D90}},
  \bibinfo{pages}{043005} (\bibinfo{year}{2014}), \eprint{1406.2160}.

\bibitem[{\citenamefont{Kowalski}(2014)}]{Kowalski:2014zda}
\bibinfo{author}{\bibfnamefont{M.}~\bibnamefont{Kowalski}}
  (\bibinfo{year}{2014}), \eprint{1411.4385}.

\bibitem[{\citenamefont{Murase}(2014)}]{Murase:2014JSI}
\bibinfo{author}{\bibfnamefont{K.}~\bibnamefont{Murase}}, \bibinfo{journal}{in
  talks presented at the 2014 JSI Workshop on Multimessenger Astronomy in the
  Era of PeV Neutrinos}  (\bibinfo{year}{2014}).

\bibitem[{\citenamefont{Murase and Waxman}(2015)}]{Murase:2015ipa}
\bibinfo{author}{\bibfnamefont{K.}~\bibnamefont{Murase}} \bibnamefont{and}
  \bibinfo{author}{\bibfnamefont{E.}~\bibnamefont{Waxman}},
  \bibinfo{journal}{in talks presented at the IPA Symposium 2015}
  (\bibinfo{year}{2015}).

\bibitem[{\citenamefont{Ackermann et~al.}(2016)}]{TheFermi-LAT:2015ykq}
\bibinfo{author}{\bibfnamefont{M.}~\bibnamefont{Ackermann}}
  \bibnamefont{et~al.} (\bibinfo{collaboration}{Fermi LAT Collaboration}),
  \bibinfo{journal}{Phys. Rev. Lett.} \textbf{\bibinfo{volume}{116}},
  \bibinfo{pages}{151105} (\bibinfo{year}{2016}), \eprint{1511.00693}.

\bibitem[{\citenamefont{Lisanti et~al.}(2016)\citenamefont{Lisanti,
  Mishra-Sharma, Necib, and Safdi}}]{Lisanti:2016jub}
\bibinfo{author}{\bibfnamefont{M.}~\bibnamefont{Lisanti}},
  \bibinfo{author}{\bibfnamefont{S.}~\bibnamefont{Mishra-Sharma}},
  \bibinfo{author}{\bibfnamefont{L.}~\bibnamefont{Necib}}, \bibnamefont{and}
  \bibinfo{author}{\bibfnamefont{B.~R.} \bibnamefont{Safdi}}
  (\bibinfo{year}{2016}), \eprint{1606.04101}.

\bibitem[{\citenamefont{Thompson et~al.}(2006)\citenamefont{Thompson, Quataert,
  Waxman, and Loeb}}]{Thompson:2006np}
\bibinfo{author}{\bibfnamefont{T.~A.} \bibnamefont{Thompson}},
  \bibinfo{author}{\bibfnamefont{E.}~\bibnamefont{Quataert}},
  \bibinfo{author}{\bibfnamefont{E.}~\bibnamefont{Waxman}}, \bibnamefont{and}
  \bibinfo{author}{\bibfnamefont{A.}~\bibnamefont{Loeb}}
  (\bibinfo{year}{2006}), \eprint{astro-ph/0608699}.

\bibitem[{\citenamefont{Lacki et~al.}(2011)\citenamefont{Lacki, Thompson,
  Quataert, Loeb, and Waxman}}]{Lacki:2010vs}
\bibinfo{author}{\bibfnamefont{B.~C.} \bibnamefont{Lacki}},
  \bibinfo{author}{\bibfnamefont{T.~A.} \bibnamefont{Thompson}},
  \bibinfo{author}{\bibfnamefont{E.}~\bibnamefont{Quataert}},
  \bibinfo{author}{\bibfnamefont{A.}~\bibnamefont{Loeb}}, \bibnamefont{and}
  \bibinfo{author}{\bibfnamefont{E.}~\bibnamefont{Waxman}},
  \bibinfo{journal}{Astrophys. J.} \textbf{\bibinfo{volume}{734}},
  \bibinfo{pages}{107} (\bibinfo{year}{2011}), \eprint{1003.3257}.

\bibitem[{\citenamefont{Lacki et~al.}(2014)\citenamefont{Lacki, Horiuchi, and
  Beacom}}]{Lacki:2012si}
\bibinfo{author}{\bibfnamefont{B.~C.} \bibnamefont{Lacki}},
  \bibinfo{author}{\bibfnamefont{S.}~\bibnamefont{Horiuchi}}, \bibnamefont{and}
  \bibinfo{author}{\bibfnamefont{J.~F.} \bibnamefont{Beacom}},
  \bibinfo{journal}{Astrophys. J.} \textbf{\bibinfo{volume}{786}},
  \bibinfo{pages}{40} (\bibinfo{year}{2014}), \eprint{1206.0772}.

\bibitem[{\citenamefont{Aab et~al.}(2015)}]{ThePierreAuger:2015rha}
\bibinfo{author}{\bibfnamefont{A.}~\bibnamefont{Aab}} \bibnamefont{et~al.}
  (\bibinfo{collaboration}{Pierre Auger Collaboration}),
  \bibinfo{journal}{JCAP} \textbf{\bibinfo{volume}{1508}}, \bibinfo{pages}{049}
  (\bibinfo{year}{2015}), \eprint{1503.07786}.

\bibitem[{Aab(2015)}]{Aab:2015bza}
\emph{\bibinfo{title}{{The Pierre Auger Observatory: Contributions to the 34th
  International Cosmic Ray Conference (ICRC 2015)}}} (\bibinfo{year}{2015}),
  \eprint{1509.03732}.

\bibitem[{\citenamefont{Decerprit and Allard}(2011)}]{Decerprit:2011qe}
\bibinfo{author}{\bibfnamefont{G.}~\bibnamefont{Decerprit}} \bibnamefont{and}
  \bibinfo{author}{\bibfnamefont{D.}~\bibnamefont{Allard}},
  \bibinfo{journal}{Astron. Astrophys.} \textbf{\bibinfo{volume}{535}},
  \bibinfo{pages}{A66} (\bibinfo{year}{2011}), \eprint{1107.3722}.

\bibitem[{\citenamefont{Aartsen et~al.}(2016{\natexlab{a}})}]{Aartsen:2016xlq}
\bibinfo{author}{\bibfnamefont{M.~G.} \bibnamefont{Aartsen}}
  \bibnamefont{et~al.} (\bibinfo{collaboration}{IceCube Collaboration})
  (\bibinfo{year}{2016}{\natexlab{a}}), \eprint{1607.08006}.

\bibitem[{\citenamefont{Aartsen et~al.}(2015{\natexlab{e}})}]{Aartsen:2014ivk}
\bibinfo{author}{\bibfnamefont{M.~G.} \bibnamefont{Aartsen}}
  \bibnamefont{et~al.} (\bibinfo{collaboration}{IceCube Collaboration}),
  \bibinfo{journal}{Astropart. Phys.} \textbf{\bibinfo{volume}{66}},
  \bibinfo{pages}{39} (\bibinfo{year}{2015}{\natexlab{e}}), \eprint{1408.0634}.

\bibitem[{\citenamefont{Aartsen et~al.}(2014{\natexlab{b}})}]{Aartsen:2014cva}
\bibinfo{author}{\bibfnamefont{M.~G.} \bibnamefont{Aartsen}}
  \bibnamefont{et~al.} (\bibinfo{collaboration}{IceCube Collaboration}),
  \bibinfo{journal}{Astrophys. J.} \textbf{\bibinfo{volume}{796}},
  \bibinfo{pages}{109} (\bibinfo{year}{2014}{\natexlab{b}}),
  \eprint{1406.6757}.

\bibitem[{\citenamefont{Aartsen et~al.}(2015{\natexlab{f}})}]{Aartsen:2015yva}
\bibinfo{author}{\bibfnamefont{M.~G.} \bibnamefont{Aartsen}}
  \bibnamefont{et~al.} (\bibinfo{collaboration}{IceCube Collaboration})
  (\bibinfo{year}{2015}{\natexlab{f}}), \eprint{1510.05222}.

\bibitem[{\citenamefont{Feldman and Cousins}(1998)}]{Feldman:1997qc}
\bibinfo{author}{\bibfnamefont{G.~J.} \bibnamefont{Feldman}} \bibnamefont{and}
  \bibinfo{author}{\bibfnamefont{R.~D.} \bibnamefont{Cousins}},
  \bibinfo{journal}{Phys.Rev.} \textbf{\bibinfo{volume}{D57}},
  \bibinfo{pages}{3873} (\bibinfo{year}{1998}), \eprint{physics/9711021}.

\bibitem[{\citenamefont{Waxman and Loeb}(2009)}]{Waxman:2008bj}
\bibinfo{author}{\bibfnamefont{E.}~\bibnamefont{Waxman}} \bibnamefont{and}
  \bibinfo{author}{\bibfnamefont{A.}~\bibnamefont{Loeb}},
  \bibinfo{journal}{JCAP} \textbf{\bibinfo{volume}{0908}}, \bibinfo{pages}{026}
  (\bibinfo{year}{2009}), \eprint{0809.3788}.

\bibitem[{\citenamefont{Hopkins and Beacom}(2006)}]{Hopkins:2006bw}
\bibinfo{author}{\bibfnamefont{A.~M.} \bibnamefont{Hopkins}} \bibnamefont{and}
  \bibinfo{author}{\bibfnamefont{J.~F.} \bibnamefont{Beacom}},
  \bibinfo{journal}{Astrophys. J.} \textbf{\bibinfo{volume}{651}},
  \bibinfo{pages}{142} (\bibinfo{year}{2006}), \eprint{astro-ph/0601463}.

\bibitem[{\citenamefont{Ajello et~al.}(2014)\citenamefont{Ajello, Romani,
  Gasparrini, Shaw, Bolmer et~al.}}]{Ajello:2013lka}
\bibinfo{author}{\bibfnamefont{M.}~\bibnamefont{Ajello}},
  \bibinfo{author}{\bibfnamefont{R.}~\bibnamefont{Romani}},
  \bibinfo{author}{\bibfnamefont{D.}~\bibnamefont{Gasparrini}},
  \bibinfo{author}{\bibfnamefont{M.}~\bibnamefont{Shaw}},
  \bibinfo{author}{\bibfnamefont{J.}~\bibnamefont{Bolmer}},
  \bibnamefont{et~al.}, \bibinfo{journal}{Astrophys.J.}
  \textbf{\bibinfo{volume}{780}}, \bibinfo{pages}{73} (\bibinfo{year}{2014}),
  \eprint{1310.0006}.

\bibitem[{\citenamefont{Ueda et~al.}(2014)\citenamefont{Ueda, Akiyama,
  Hasinger, Miyaji, and Watson}}]{Ueda:2014tma}
\bibinfo{author}{\bibfnamefont{Y.}~\bibnamefont{Ueda}},
  \bibinfo{author}{\bibfnamefont{M.}~\bibnamefont{Akiyama}},
  \bibinfo{author}{\bibfnamefont{G.}~\bibnamefont{Hasinger}},
  \bibinfo{author}{\bibfnamefont{T.}~\bibnamefont{Miyaji}}, \bibnamefont{and}
  \bibinfo{author}{\bibfnamefont{M.~G.} \bibnamefont{Watson}},
  \bibinfo{journal}{Astrophys.J.} \textbf{\bibinfo{volume}{786}},
  \bibinfo{pages}{104} (\bibinfo{year}{2014}), \eprint{1402.1836}.

\bibitem[{\citenamefont{Aartsen et~al.}(2014{\natexlab{c}})}]{Aartsen:2014njl}
\bibinfo{author}{\bibfnamefont{M.}~\bibnamefont{Aartsen}} \bibnamefont{et~al.}
  (\bibinfo{collaboration}{IceCube-Gen2 Collaboration})
  (\bibinfo{year}{2014}{\natexlab{c}}), \eprint{1412.5106}.

\bibitem[{\citenamefont{Ackermann et~al.}(2012)}]{Ackermann:2012vca}
\bibinfo{author}{\bibfnamefont{M.}~\bibnamefont{Ackermann}}
  \bibnamefont{et~al.} (\bibinfo{collaboration}{Fermi LAT Collaboration}),
  \bibinfo{journal}{Astrophys.J.} \textbf{\bibinfo{volume}{755}},
  \bibinfo{pages}{164} (\bibinfo{year}{2012}), \eprint{1206.1346}.

\bibitem[{\citenamefont{Gruppioni et~al.}(2013)\citenamefont{Gruppioni, Pozzi,
  Rodighiero, Delvecchio, Berta et~al.}}]{Gruppioni:2013jna}
\bibinfo{author}{\bibfnamefont{C.}~\bibnamefont{Gruppioni}},
  \bibinfo{author}{\bibfnamefont{F.}~\bibnamefont{Pozzi}},
  \bibinfo{author}{\bibfnamefont{G.}~\bibnamefont{Rodighiero}},
  \bibinfo{author}{\bibfnamefont{I.}~\bibnamefont{Delvecchio}},
  \bibinfo{author}{\bibfnamefont{S.}~\bibnamefont{Berta}},
  \bibnamefont{et~al.}, \bibinfo{journal}{Mon.Not.Roy.Astron.Soc.}
  \textbf{\bibinfo{volume}{432}}, \bibinfo{pages}{23} (\bibinfo{year}{2013}),
  \eprint{1302.5209}.

\bibitem[{\citenamefont{Warren et~al.}(2006)\citenamefont{Warren, Abazajian,
  Holz, and Teodoro}}]{Warren:2005ey}
\bibinfo{author}{\bibfnamefont{M.~S.} \bibnamefont{Warren}},
  \bibinfo{author}{\bibfnamefont{K.}~\bibnamefont{Abazajian}},
  \bibinfo{author}{\bibfnamefont{D.~E.} \bibnamefont{Holz}}, \bibnamefont{and}
  \bibinfo{author}{\bibfnamefont{L.}~\bibnamefont{Teodoro}},
  \bibinfo{journal}{Astrophys. J.} \textbf{\bibinfo{volume}{646}},
  \bibinfo{pages}{881} (\bibinfo{year}{2006}), \eprint{astro-ph/0506395}.

\bibitem[{\citenamefont{Inoue}(2011)}]{Inoue:2011bm}
\bibinfo{author}{\bibfnamefont{Y.}~\bibnamefont{Inoue}},
  \bibinfo{journal}{Astrophys. J.} \textbf{\bibinfo{volume}{733}},
  \bibinfo{pages}{66} (\bibinfo{year}{2011}), \eprint{1103.3946}.

\bibitem[{\citenamefont{Willott et~al.}(2001)\citenamefont{Willott, Rawlings,
  Blundell, Lacy, and Eales}}]{Willott:2000dh}
\bibinfo{author}{\bibfnamefont{C.~J.} \bibnamefont{Willott}},
  \bibinfo{author}{\bibfnamefont{S.}~\bibnamefont{Rawlings}},
  \bibinfo{author}{\bibfnamefont{K.~M.} \bibnamefont{Blundell}},
  \bibinfo{author}{\bibfnamefont{M.}~\bibnamefont{Lacy}}, \bibnamefont{and}
  \bibinfo{author}{\bibfnamefont{S.~A.} \bibnamefont{Eales}},
  \bibinfo{journal}{Mon. Not. Roy. Astron. Soc.}
  \textbf{\bibinfo{volume}{322}}, \bibinfo{pages}{536} (\bibinfo{year}{2001}),
  \eprint{astro-ph/0010419}.

\bibitem[{\citenamefont{Ho}(2008)}]{Ho:2008rf}
\bibinfo{author}{\bibfnamefont{L.~C.} \bibnamefont{Ho}}, \bibinfo{journal}{Ann.
  Rev. Astron. Astrophys.} \textbf{\bibinfo{volume}{46}}, \bibinfo{pages}{475}
  (\bibinfo{year}{2008}), \eprint{0803.2268}.

\bibitem[{\citenamefont{Tamborra et~al.}(2014)\citenamefont{Tamborra, Ando, and
  Murase}}]{Tamborra:2014xia}
\bibinfo{author}{\bibfnamefont{I.}~\bibnamefont{Tamborra}},
  \bibinfo{author}{\bibfnamefont{S.}~\bibnamefont{Ando}}, \bibnamefont{and}
  \bibinfo{author}{\bibfnamefont{K.}~\bibnamefont{Murase}},
  \bibinfo{journal}{JCAP} \textbf{\bibinfo{volume}{1409}}, \bibinfo{pages}{043}
  (\bibinfo{year}{2014}), \eprint{1404.1189}.

\bibitem[{\citenamefont{Wang and Loeb}(2016)}]{Wang:2016vbf}
\bibinfo{author}{\bibfnamefont{X.}~\bibnamefont{Wang}} \bibnamefont{and}
  \bibinfo{author}{\bibfnamefont{A.}~\bibnamefont{Loeb}}
  (\bibinfo{year}{2016}), \eprint{1607.06476}.

\bibitem[{\citenamefont{Keshet et~al.}(2003)\citenamefont{Keshet, Waxman, Loeb,
  Springel, and Hernquist}}]{Keshet:2002sw}
\bibinfo{author}{\bibfnamefont{U.}~\bibnamefont{Keshet}},
  \bibinfo{author}{\bibfnamefont{E.}~\bibnamefont{Waxman}},
  \bibinfo{author}{\bibfnamefont{A.}~\bibnamefont{Loeb}},
  \bibinfo{author}{\bibfnamefont{V.}~\bibnamefont{Springel}}, \bibnamefont{and}
  \bibinfo{author}{\bibfnamefont{L.}~\bibnamefont{Hernquist}},
  \bibinfo{journal}{Astrophys. J.} \textbf{\bibinfo{volume}{585}},
  \bibinfo{pages}{128} (\bibinfo{year}{2003}), \eprint{astro-ph/0202318}.

\bibitem[{\citenamefont{Kushnir and Waxman}(2009)}]{Kushnir:2009vm}
\bibinfo{author}{\bibfnamefont{D.}~\bibnamefont{Kushnir}} \bibnamefont{and}
  \bibinfo{author}{\bibfnamefont{E.}~\bibnamefont{Waxman}},
  \bibinfo{journal}{JCAP} \textbf{\bibinfo{volume}{0908}}, \bibinfo{pages}{002}
  (\bibinfo{year}{2009}), \eprint{0903.2271}.

\bibitem[{\citenamefont{Colafrancesco and Blasi}(1998)}]{Colafrancesco:1998us}
\bibinfo{author}{\bibfnamefont{S.}~\bibnamefont{Colafrancesco}}
  \bibnamefont{and} \bibinfo{author}{\bibfnamefont{P.}~\bibnamefont{Blasi}},
  \bibinfo{journal}{Astropart.Phys.} \textbf{\bibinfo{volume}{9}},
  \bibinfo{pages}{227} (\bibinfo{year}{1998}), \eprint{astro-ph/9804262}.

\bibitem[{\citenamefont{Reiprich and Boehringer}(2002)}]{Reiprich:2001zv}
\bibinfo{author}{\bibfnamefont{T.~H.} \bibnamefont{Reiprich}} \bibnamefont{and}
  \bibinfo{author}{\bibfnamefont{H.}~\bibnamefont{Boehringer}},
  \bibinfo{journal}{Astrophys. J.} \textbf{\bibinfo{volume}{567}},
  \bibinfo{pages}{716} (\bibinfo{year}{2002}), \eprint{astro-ph/0111285}.

\bibitem[{\citenamefont{Berezinsky et~al.}(1997)\citenamefont{Berezinsky,
  Blasi, and Ptuskin}}]{Berezinsky:1996wx}
\bibinfo{author}{\bibfnamefont{V.}~\bibnamefont{Berezinsky}},
  \bibinfo{author}{\bibfnamefont{P.}~\bibnamefont{Blasi}}, \bibnamefont{and}
  \bibinfo{author}{\bibfnamefont{V.}~\bibnamefont{Ptuskin}},
  \bibinfo{journal}{Astrophys J.} \textbf{\bibinfo{volume}{487}},
  \bibinfo{pages}{529} (\bibinfo{year}{1997}), \eprint{astro-ph/9609048}.

\bibitem[{\citenamefont{Zandanel et~al.}(2015)\citenamefont{Zandanel, Tamborra,
  Gabici, and Ando}}]{Zandanel:2014pva}
\bibinfo{author}{\bibfnamefont{F.}~\bibnamefont{Zandanel}},
  \bibinfo{author}{\bibfnamefont{I.}~\bibnamefont{Tamborra}},
  \bibinfo{author}{\bibfnamefont{S.}~\bibnamefont{Gabici}}, \bibnamefont{and}
  \bibinfo{author}{\bibfnamefont{S.}~\bibnamefont{Ando}},
  \bibinfo{journal}{Astron. Astrophys.} \textbf{\bibinfo{volume}{578}},
  \bibinfo{pages}{A32} (\bibinfo{year}{2015}), \eprint{1410.8697}.

\bibitem[{\citenamefont{Fang and Olinto}(2016)}]{Fang:2016amf}
\bibinfo{author}{\bibfnamefont{K.}~\bibnamefont{Fang}} \bibnamefont{and}
  \bibinfo{author}{\bibfnamefont{A.~V.} \bibnamefont{Olinto}},
  \bibinfo{journal}{Astrophys. J.} \textbf{\bibinfo{volume}{828}},
  \bibinfo{pages}{37} (\bibinfo{year}{2016}), \eprint{1607.00380}.

\bibitem[{\citenamefont{Acero et~al.}(2015)}]{Acero:2015hja}
\bibinfo{author}{\bibfnamefont{F.}~\bibnamefont{Acero}} \bibnamefont{et~al.}
  (\bibinfo{collaboration}{Fermi LAT Collaboration}),
  \bibinfo{journal}{Astrophys.J.Suppl.} \textbf{\bibinfo{volume}{218}},
  \bibinfo{pages}{23} (\bibinfo{year}{2015}), \eprint{1501.02003}.

\bibitem[{\citenamefont{Pfrommer}(2013)}]{Pfrommer:2013eoa}
\bibinfo{author}{\bibfnamefont{C.}~\bibnamefont{Pfrommer}},
  \bibinfo{journal}{Astrophys. J.} \textbf{\bibinfo{volume}{779}},
  \bibinfo{pages}{10} (\bibinfo{year}{2013}), \eprint{1303.5443}.

\bibitem[{\citenamefont{Fujita et~al.}(2015)\citenamefont{Fujita, Kimura, and
  Murase}}]{Fujita:2015xva}
\bibinfo{author}{\bibfnamefont{Y.}~\bibnamefont{Fujita}},
  \bibinfo{author}{\bibfnamefont{S.~S.} \bibnamefont{Kimura}},
  \bibnamefont{and} \bibinfo{author}{\bibfnamefont{K.}~\bibnamefont{Murase}},
  \bibinfo{journal}{Phys. Rev.} \textbf{\bibinfo{volume}{D92}},
  \bibinfo{pages}{023001} (\bibinfo{year}{2015}), \eprint{1506.05461}.

\bibitem[{\citenamefont{Alvarez-Muniz and
  M\'esz\'aros}(2004)}]{AlvarezMuniz:2004uz}
\bibinfo{author}{\bibfnamefont{J.}~\bibnamefont{Alvarez-Muniz}}
  \bibnamefont{and}
  \bibinfo{author}{\bibfnamefont{P.}~\bibnamefont{M\'esz\'aros}},
  \bibinfo{journal}{Phys.Rev.} \textbf{\bibinfo{volume}{D70}},
  \bibinfo{pages}{123001} (\bibinfo{year}{2004}), \eprint{astro-ph/0409034}.

\bibitem[{\citenamefont{Anchordoqui et~al.}(2014)\citenamefont{Anchordoqui,
  Paul, da~Silva, Torres, and Vlcek}}]{Anchordoqui:2014yva}
\bibinfo{author}{\bibfnamefont{L.~A.} \bibnamefont{Anchordoqui}},
  \bibinfo{author}{\bibfnamefont{T.~C.} \bibnamefont{Paul}},
  \bibinfo{author}{\bibfnamefont{L.~H.~M.} \bibnamefont{da~Silva}},
  \bibinfo{author}{\bibfnamefont{D.~F.} \bibnamefont{Torres}},
  \bibnamefont{and} \bibinfo{author}{\bibfnamefont{B.~J.} \bibnamefont{Vlcek}},
  \bibinfo{journal}{Phys.Rev.} \textbf{\bibinfo{volume}{D89}},
  \bibinfo{pages}{127304} (\bibinfo{year}{2014}), \eprint{1405.7648}.

\bibitem[{\citenamefont{Padovani and Resconi}(2014)}]{Padovani:2014bha}
\bibinfo{author}{\bibfnamefont{P.}~\bibnamefont{Padovani}} \bibnamefont{and}
  \bibinfo{author}{\bibfnamefont{E.}~\bibnamefont{Resconi}},
  \bibinfo{journal}{Mon.Not.Roy.Astron.Soc.} \textbf{\bibinfo{volume}{443}},
  \bibinfo{pages}{474} (\bibinfo{year}{2014}), \eprint{1406.0376}.

\bibitem[{\citenamefont{Sahu and Miranda}(2015)}]{Sahu:2014fua}
\bibinfo{author}{\bibfnamefont{S.}~\bibnamefont{Sahu}} \bibnamefont{and}
  \bibinfo{author}{\bibfnamefont{L.~S.} \bibnamefont{Miranda}},
  \bibinfo{journal}{Eur. Phys. J.} \textbf{\bibinfo{volume}{C75}},
  \bibinfo{pages}{273} (\bibinfo{year}{2015}), \eprint{1408.3664}.

\bibitem[{\citenamefont{Moharana and Razzaque}(2015)}]{Moharana:2015nxa}
\bibinfo{author}{\bibfnamefont{R.}~\bibnamefont{Moharana}} \bibnamefont{and}
  \bibinfo{author}{\bibfnamefont{S.}~\bibnamefont{Razzaque}},
  \bibinfo{journal}{JCAP} \textbf{\bibinfo{volume}{1508}}, \bibinfo{pages}{014}
  (\bibinfo{year}{2015}), \eprint{1501.05158}.

\bibitem[{\citenamefont{Emig et~al.}(2015)\citenamefont{Emig, Lunardini, and
  Windhorst}}]{Emig:2015dma}
\bibinfo{author}{\bibfnamefont{K.}~\bibnamefont{Emig}},
  \bibinfo{author}{\bibfnamefont{C.}~\bibnamefont{Lunardini}},
  \bibnamefont{and}
  \bibinfo{author}{\bibfnamefont{R.}~\bibnamefont{Windhorst}},
  \bibinfo{journal}{JCAP} \textbf{\bibinfo{volume}{1512}}, \bibinfo{pages}{029}
  (\bibinfo{year}{2015}), \eprint{1507.05711}.

\bibitem[{\citenamefont{Moharana and Razzaque}(2016)}]{Moharana:2016mkl}
\bibinfo{author}{\bibfnamefont{R.}~\bibnamefont{Moharana}} \bibnamefont{and}
  \bibinfo{author}{\bibfnamefont{S.}~\bibnamefont{Razzaque}}
  (\bibinfo{year}{2016}), \eprint{1606.04420}.

\bibitem[{\citenamefont{Wang and Li}(2016)}]{Wang:2015woa}
\bibinfo{author}{\bibfnamefont{B.}~\bibnamefont{Wang}} \bibnamefont{and}
  \bibinfo{author}{\bibfnamefont{Z.}~\bibnamefont{Li}}, \bibinfo{journal}{Sci.
  China Phys. Mech. Astron.} \textbf{\bibinfo{volume}{59}},
  \bibinfo{pages}{619502} (\bibinfo{year}{2016}), \eprint{1505.04418}.

\bibitem[{\citenamefont{Gl$\ddot{\rm u}$senkamp}(2015)}]{Glusenkamp:2015jca}
\bibinfo{author}{\bibfnamefont{T.}~\bibnamefont{Gl$\ddot{\rm u}$senkamp}}
  (\bibinfo{collaboration}{IceCube Collaboration}) (\bibinfo{year}{2015}),
  \eprint{1502.03104}.

\bibitem[{\citenamefont{Aartsen et~al.}(2016{\natexlab{b}})}]{Aartsen:2016oji}
\bibinfo{author}{\bibfnamefont{M.~G.} \bibnamefont{Aartsen}}
  \bibnamefont{et~al.} (\bibinfo{collaboration}{IceCube Collaboration})
  (\bibinfo{year}{2016}{\natexlab{b}}), \eprint{1609.04981}.

\bibitem[{\citenamefont{Tang et~al.}(2014)\citenamefont{Tang, Wang, and
  Thomas~Tam}}]{Tang:2014dia}
\bibinfo{author}{\bibfnamefont{Q.-W.} \bibnamefont{Tang}},
  \bibinfo{author}{\bibfnamefont{X.-Y.} \bibnamefont{Wang}}, \bibnamefont{and}
  \bibinfo{author}{\bibfnamefont{P.-H.} \bibnamefont{Thomas~Tam}},
  \bibinfo{journal}{Astrophys.J.} \textbf{\bibinfo{volume}{794}},
  \bibinfo{pages}{26} (\bibinfo{year}{2014}), \eprint{1407.3391}.

\bibitem[{\citenamefont{Peng et~al.}(2016)\citenamefont{Peng, Wang, Liu, Tang,
  and Wang}}]{Peng:2016nsx}
\bibinfo{author}{\bibfnamefont{F.-K.} \bibnamefont{Peng}},
  \bibinfo{author}{\bibfnamefont{X.-Y.} \bibnamefont{Wang}},
  \bibinfo{author}{\bibfnamefont{R.-Y.} \bibnamefont{Liu}},
  \bibinfo{author}{\bibfnamefont{Q.-W.} \bibnamefont{Tang}}, \bibnamefont{and}
  \bibinfo{author}{\bibfnamefont{J.-F.} \bibnamefont{Wang}},
  \bibinfo{journal}{Astrophys. J.} \textbf{\bibinfo{volume}{821}},
  \bibinfo{pages}{L20} (\bibinfo{year}{2016}), \eprint{1603.06355}.

\bibitem[{\citenamefont{Griffin et~al.}(2016)\citenamefont{Griffin, Dai, and
  Thompson}}]{Griffin:2016wzb}
\bibinfo{author}{\bibfnamefont{R.~D.} \bibnamefont{Griffin}},
  \bibinfo{author}{\bibfnamefont{X.}~\bibnamefont{Dai}}, \bibnamefont{and}
  \bibinfo{author}{\bibfnamefont{T.~A.} \bibnamefont{Thompson}},
  \bibinfo{journal}{Astrophys. J.} \textbf{\bibinfo{volume}{823}},
  \bibinfo{pages}{L17} (\bibinfo{year}{2016}), \eprint{1603.06949}.

\bibitem[{\citenamefont{Murase and Beacom}(2013)}]{Murase:2012rd}
\bibinfo{author}{\bibfnamefont{K.}~\bibnamefont{Murase}} \bibnamefont{and}
  \bibinfo{author}{\bibfnamefont{J.~F.} \bibnamefont{Beacom}},
  \bibinfo{journal}{JCAP} \textbf{\bibinfo{volume}{1302}}, \bibinfo{pages}{028}
  (\bibinfo{year}{2013}), \eprint{1209.0225}.

\bibitem[{\citenamefont{Kneiske et~al.}(2004)\citenamefont{Kneiske, Bretz,
  Mannheim, and Hartmann}}]{Kneiske:2003tx}
\bibinfo{author}{\bibfnamefont{T.~M.} \bibnamefont{Kneiske}},
  \bibinfo{author}{\bibfnamefont{T.}~\bibnamefont{Bretz}},
  \bibinfo{author}{\bibfnamefont{K.}~\bibnamefont{Mannheim}}, \bibnamefont{and}
  \bibinfo{author}{\bibfnamefont{D.~H.} \bibnamefont{Hartmann}},
  \bibinfo{journal}{Astron. Astrophys.} \textbf{\bibinfo{volume}{413}},
  \bibinfo{pages}{807} (\bibinfo{year}{2004}), \eprint{astro-ph/0309141}.

\bibitem[{\citenamefont{Actis et~al.}(2011)}]{Consortium:2010bc}
\bibinfo{author}{\bibfnamefont{M.}~\bibnamefont{Actis}} \bibnamefont{et~al.}
  (\bibinfo{collaboration}{CTA Consortium}), \bibinfo{journal}{Exper.Astron.}
  \textbf{\bibinfo{volume}{32}}, \bibinfo{pages}{193} (\bibinfo{year}{2011}),
  \eprint{1008.3703}.

\bibitem[{\citenamefont{Abeysekara et~al.}(2013)\citenamefont{Abeysekara,
  Alfaro, Alvarez, Álvarez, Arceo et~al.}}]{Abeysekara:2013tza}
\bibinfo{author}{\bibfnamefont{A.}~\bibnamefont{Abeysekara}},
  \bibinfo{author}{\bibfnamefont{R.}~\bibnamefont{Alfaro}},
  \bibinfo{author}{\bibfnamefont{C.}~\bibnamefont{Alvarez}},
  \bibinfo{author}{\bibfnamefont{J.}~\bibnamefont{Álvarez}},
  \bibinfo{author}{\bibfnamefont{R.}~\bibnamefont{Arceo}},
  \bibnamefont{et~al.}, \bibinfo{journal}{Astropart.Phys.}
  \textbf{\bibinfo{volume}{50-52}}, \bibinfo{pages}{26} (\bibinfo{year}{2013}),
  \eprint{1306.5800}.

\bibitem[{\citenamefont{Cuoco et~al.}(2012)\citenamefont{Cuoco, Komatsu, and
  Siegal-Gaskins}}]{Cuoco:2012yf}
\bibinfo{author}{\bibfnamefont{A.}~\bibnamefont{Cuoco}},
  \bibinfo{author}{\bibfnamefont{E.}~\bibnamefont{Komatsu}}, \bibnamefont{and}
  \bibinfo{author}{\bibfnamefont{J.~M.} \bibnamefont{Siegal-Gaskins}},
  \bibinfo{journal}{Phys.Rev.} \textbf{\bibinfo{volume}{D86}},
  \bibinfo{pages}{063004} (\bibinfo{year}{2012}), \eprint{1202.5309}.

\bibitem[{\citenamefont{Zechlin et~al.}(2015)\citenamefont{Zechlin, Cuoco,
  Donato, Fornengo, and Vittino}}]{Zechlin:2015wdz}
\bibinfo{author}{\bibfnamefont{H.-S.} \bibnamefont{Zechlin}},
  \bibinfo{author}{\bibfnamefont{A.}~\bibnamefont{Cuoco}},
  \bibinfo{author}{\bibfnamefont{F.}~\bibnamefont{Donato}},
  \bibinfo{author}{\bibfnamefont{N.}~\bibnamefont{Fornengo}}, \bibnamefont{and}
  \bibinfo{author}{\bibfnamefont{A.}~\bibnamefont{Vittino}}
  (\bibinfo{year}{2015}), \eprint{1512.07190}.

\bibitem[{\citenamefont{Zechlin et~al.}(2016)\citenamefont{Zechlin, Cuoco,
  Donato, Fornengo, and Regis}}]{Zechlin:2016pme}
\bibinfo{author}{\bibfnamefont{H.-S.} \bibnamefont{Zechlin}},
  \bibinfo{author}{\bibfnamefont{A.}~\bibnamefont{Cuoco}},
  \bibinfo{author}{\bibfnamefont{F.}~\bibnamefont{Donato}},
  \bibinfo{author}{\bibfnamefont{N.}~\bibnamefont{Fornengo}}, \bibnamefont{and}
  \bibinfo{author}{\bibfnamefont{M.}~\bibnamefont{Regis}}
  (\bibinfo{year}{2016}), \eprint{1605.04256}.

\bibitem[{\citenamefont{Ando et~al.}(2015)\citenamefont{Ando, Tamborra, and
  Zandanel}}]{Ando:2015bva}
\bibinfo{author}{\bibfnamefont{S.}~\bibnamefont{Ando}},
  \bibinfo{author}{\bibfnamefont{I.}~\bibnamefont{Tamborra}}, \bibnamefont{and}
  \bibinfo{author}{\bibfnamefont{F.}~\bibnamefont{Zandanel}},
  \bibinfo{journal}{Phys. Rev. Lett.} \textbf{\bibinfo{volume}{115}},
  \bibinfo{pages}{221101} (\bibinfo{year}{2015}), \eprint{1509.02444}.

\bibitem[{\citenamefont{Petropoulou et~al.}(2016)\citenamefont{Petropoulou,
  Coenders, and Dimitrakoudis}}]{Petropoulou:2016ujj}
\bibinfo{author}{\bibfnamefont{M.}~\bibnamefont{Petropoulou}},
  \bibinfo{author}{\bibfnamefont{S.}~\bibnamefont{Coenders}}, \bibnamefont{and}
  \bibinfo{author}{\bibfnamefont{S.}~\bibnamefont{Dimitrakoudis}},
  \bibinfo{journal}{Astropart. Phys.} \textbf{\bibinfo{volume}{80}},
  \bibinfo{pages}{115} (\bibinfo{year}{2016}), \eprint{1603.06954}.

\bibitem[{\citenamefont{Kadler et~al.}(2016)}]{Kadler:2016ygj}
\bibinfo{author}{\bibfnamefont{M.}~\bibnamefont{Kadler}} \bibnamefont{et~al.},
  \bibinfo{journal}{Nature Phys.} \textbf{\bibinfo{volume}{12}},
  \bibinfo{pages}{807} (\bibinfo{year}{2016}), \eprint{1602.02012}.

\bibitem[{\citenamefont{Laha et~al.}(2013)\citenamefont{Laha, Beacom, Dasgupta,
  Horiuchi, and Murase}}]{Laha:2013lka}
\bibinfo{author}{\bibfnamefont{R.}~\bibnamefont{Laha}},
  \bibinfo{author}{\bibfnamefont{J.~F.} \bibnamefont{Beacom}},
  \bibinfo{author}{\bibfnamefont{B.}~\bibnamefont{Dasgupta}},
  \bibinfo{author}{\bibfnamefont{S.}~\bibnamefont{Horiuchi}}, \bibnamefont{and}
  \bibinfo{author}{\bibfnamefont{K.}~\bibnamefont{Murase}},
  \bibinfo{journal}{Phys.Rev.} \textbf{\bibinfo{volume}{D88}},
  \bibinfo{pages}{043009} (\bibinfo{year}{2013}), \eprint{1306.2309}.

\bibitem[{\citenamefont{Blum et~al.}(2014)\citenamefont{Blum, Hook, and
  Murase}}]{Blum:2014ewa}
\bibinfo{author}{\bibfnamefont{K.}~\bibnamefont{Blum}},
  \bibinfo{author}{\bibfnamefont{A.}~\bibnamefont{Hook}}, \bibnamefont{and}
  \bibinfo{author}{\bibfnamefont{K.}~\bibnamefont{Murase}}
  (\bibinfo{year}{2014}), \eprint{1408.3799}.

\bibitem[{\citenamefont{Connolly et~al.}(2011)\citenamefont{Connolly, Thorne,
  and Waters}}]{Connolly:2011vc}
\bibinfo{author}{\bibfnamefont{A.}~\bibnamefont{Connolly}},
  \bibinfo{author}{\bibfnamefont{R.~S.} \bibnamefont{Thorne}},
  \bibnamefont{and} \bibinfo{author}{\bibfnamefont{D.}~\bibnamefont{Waters}},
  \bibinfo{journal}{Phys.Rev.} \textbf{\bibinfo{volume}{D83}},
  \bibinfo{pages}{113009} (\bibinfo{year}{2011}), \eprint{1102.0691}.

\bibitem[{\citenamefont{Dziewonski and Anderson}(1981)}]{Dziewonski:1981xy}
\bibinfo{author}{\bibfnamefont{A.}~\bibnamefont{Dziewonski}} \bibnamefont{and}
  \bibinfo{author}{\bibfnamefont{D.}~\bibnamefont{Anderson}},
  \bibinfo{journal}{Phys.Earth Planet.Interiors} \textbf{\bibinfo{volume}{25}},
  \bibinfo{pages}{297} (\bibinfo{year}{1981}).

\bibitem[{\citenamefont{Gonzalez-Garcia
  et~al.}(2009)\citenamefont{Gonzalez-Garcia, Halzen, and
  Mohapatra}}]{GonzalezGarcia:2009jc}
\bibinfo{author}{\bibfnamefont{M.}~\bibnamefont{Gonzalez-Garcia}},
  \bibinfo{author}{\bibfnamefont{F.}~\bibnamefont{Halzen}}, \bibnamefont{and}
  \bibinfo{author}{\bibfnamefont{S.}~\bibnamefont{Mohapatra}},
  \bibinfo{journal}{Astropart.Phys.} \textbf{\bibinfo{volume}{31}},
  \bibinfo{pages}{437} (\bibinfo{year}{2009}), \eprint{0902.1176}.

\bibitem[{\citenamefont{Abbasi et~al.}(2011)}]{Abbasi:2010ie}
\bibinfo{author}{\bibfnamefont{R.}~\bibnamefont{Abbasi}} \bibnamefont{et~al.}
  (\bibinfo{collaboration}{IceCube Collaboration}),
  \bibinfo{journal}{Phys.Rev.} \textbf{\bibinfo{volume}{D83}},
  \bibinfo{pages}{012001} (\bibinfo{year}{2011}), \eprint{1010.3980}.

\bibitem[{\citenamefont{Aartsen et~al.}(2013{\natexlab{c}})}]{Aartsen:2012uu}
\bibinfo{author}{\bibfnamefont{M.}~\bibnamefont{Aartsen}} \bibnamefont{et~al.}
  (\bibinfo{collaboration}{IceCube Collaboration}),
  \bibinfo{journal}{Phys.Rev.Lett.} \textbf{\bibinfo{volume}{110}},
  \bibinfo{pages}{151105} (\bibinfo{year}{2013}{\natexlab{c}}),
  \eprint{1212.4760}.

\end{thebibliography}




\appendix
\section{Calculating the number of through-going muon events}
Following Refs.~\cite{Laha:2013lka,Blum:2014ewa}, we calculate the differential detection rate of through-going muon tracks
\begin{eqnarray}
\left(\frac{d\dot{\mathcal N}_\mu}{dE_\mu}\right)\approx\frac{N_A{\mathcal A}_{\rm det}^{\rm IC}}{\alpha+\beta E_\mu}
\int_{E_\mu}^{\infty}dE_{\nu}\,\phi_{\nu_\mu}\sigma_{\rm CC} {\rm e}^{-\tau_{\nu N}},
\end{eqnarray}
where $E_\nu$ is the incoming neutrino energy, $E_\mu$ is the muon energy, $N_A$ is the Avogadro number, ${\mathcal A}_{\rm det}^{\rm IC}$ is the muon effective area, $\sigma_{\rm CC}$ is the charged-current cross section, and $\tau_{\nu N}$ is the attenuation in the Earth.  We use the cross sections given in Ref.~\cite{Connolly:2011vc}, the Earth model of Ref.~\cite{Dziewonski:1981xy}, and the zenith-angle dependence of ${\mathcal A}_{\rm det}^{\rm IC}$ given in Ref.~\cite{GonzalezGarcia:2009jc}.  We use an average muon energy-loss rate, $-dE_\mu/dX=\alpha+\beta E_\mu$ with $\alpha=2\times{10}^{-3}~{\rm GeV}~{\rm cm}^2~{\rm g}^{-1}$ and $\beta=4\times{10}^{-6}~{\rm cm}^2~{\rm g}^{-1}$. We have verified that the muon neutrino effective area of IceCube-86 reported in Ref.~\cite{Aartsen:2014cva} is reproduced by our calculation.

To evaluate the background, we consider both the conventional and the prompt atmospheric muon neutrino backgrounds~\cite{Abbasi:2010ie,Aartsen:2012uu}.  For example, the number of total background events (${\mathcal N}_b$) with $E_\mu\gtrsim50$~TeV in the six-year observation by IceCube is ${\mathcal N}_b\sim200$, which is consistent with Ref.~\cite{Aartsen:2016xlq}.  
In addition, we take into account the cumulative astrophysical background based on the diffuse muon neutrino flux in the northern sky~\cite{Aartsen:2015rwa}.
For IceCube, we set the angular size to $0.5~{\rm deg}+1.0~{\rm deg}~{(E_\nu/{\rm TeV})}^{-0.5}$~\cite{Aartsen:2015yva}. For {\it IceCube-Gen2}, the best angular resolution may be ${\Delta\theta}_{\rm res}=0.1$~deg~\cite{Aartsen:2014njl}.  With the kinematic limit, we use ${\Delta\theta}_{\rm res}+0.5~{\rm deg}~{(E_\nu/{\rm TeV})}^{-0.5}$.

For a given background, the limits on the single source flux are estimated following Ref.~\cite{Feldman:1997qc}.  Note that the $5\sigma$ sensitivity (discovery potential) is about four times worse than the 90\% CL limit sensitivity~\cite{Aartsen:2014cva}.  We consider only high-energy muons, for which background events within the angular resolution are essentially negligible so that the simple analysis described in the text is also applicable.  To derive muon neutrino constraints, we set the muon energy threshold to 50~TeV.  We do not consider starting muon events or neutrino-induced showers.  The constraints can be improved by including such events.  In this sense, our results are conservative.

As noted in the text, there may be false multiplet sources (with $N_b$) due to the atmospheric neutrino background. Similarly to the well-known birthday problem, for the number of angular bins, $n$, in the northern sky, the probability to find false multiplet sources is given by $p_{\geq2}=1-{}_n{\rm C}_{{\mathcal N}_b}{\mathcal N}_b!/n^{{\mathcal N}_b}$. 
The expectation value of the number of background pairs is ${}_{{\mathcal N}_b}{\rm C}_2/n$.  For example, with ${\mathcal N}_b=200$ and $n=26262$ (corresponding to ${\Delta\theta}_{\rm res}=0.5$~deg), we have $N_b\sim {}_{200}{\rm C}_2/26262\sim0.8$. 
Note that the probability to find false triplet or higher multiplet sources is given by $p_{\geq3}=1-{}_n{\rm C}_{{\mathcal N}_b}{\mathcal N}_b!/n^{{\mathcal N}_b}-\Sigma_{k=1}^{[{\mathcal N}_b/2]}({}_{{\mathcal N}_b}{\rm C}_2\cdots{}_{{\mathcal N}_b-2k+2}{\rm C}_2/k!)({}_n{\rm C}_{{\mathcal N}_b-k}({\mathcal N}_b-k)!/n^{{\mathcal N}_b})$,  
which is negligibly small in the high-energy track data.

For the signal, motivated by CR reservoir models, we first consider~\cite{Murase:2015xka}
\begin{equation}\label{eq:pp}
E_{\nu}L_{E_\nu}\propto
\begin{cases}
E_\nu^{2-s}
& (E_\nu\leq E_\nu^b)\\
E_\nu^{2-s'}
& (E_\nu^b<E_\nu)\\
\end{cases}
\,,
\end{equation}
where $E_\nu^b$ is the neutrino break energy.  Note that in the CR reservoirs such as SBGs and GCs/GGs, a spectral break around a few PeV energies due to CR diffusion is predicted~\cite{Loeb:2006tw,Murase:2008yt}.  The softening of the spectrum, $\delta\equiv s'-s$, comes from the the energy dependence of the diffusion tensor.
Throughout this work, we mainly use $s=2.0$ and $s'=2.5$ as invoked by Ref.~\cite{Murase:2013rfa}, which allow the CR reservoir models to explain the high-energy IceCube data without contradicting the diffuse gamma-ray background.  The normalization of $E_{\nu}L_{E_{\nu_\mu}}$ is set by $E_\nu^2\Phi_{\nu_\mu}={10}^{-8}~{\rm GeV}~{\rm cm}^{-2}~{\rm s}^{-1}~{\rm sr}^{-1}$ at $E_\nu=0.3$~PeV.
Note that larger indices of $s>2$ lead to larger values of $(d\dot{\mathcal N}_\mu/dE_\mu)$, leading to stronger muon neutrino limits. 
\end{document}